\documentclass[12pt,preprint]{aastex}

\def\lesssim{\mathrel{\hbox{\rlap{\hbox{\lower4pt\hbox{$\sim$}}}\hbox{$<$}}}}
\def\gtrsim{\mathrel{\hbox{\rlap{\hbox{\lower4pt\hbox{$\sim$}}}\hbox{$>$}}}}

\usepackage{graphicx}

\slugcomment{}

\title{Surface Structure in an Accretion Disk Annulus with
Comparable Radiation and Gas Pressure}

\shortauthors{Blaes, Hirose \& Krolik}

\begin{document}

\shorttitle{Surface Structure of Accretion Disks}

\author{Omer Blaes}
\affil{Department of Physics, University of California, Santa Barbara,
Santa Barbara CA 93106}

\author{Shigenobu Hirose}
\affil{The Earth Simulator Center, JAMSTEC, Yokohama, Kanagawa 236-0001, Japan}

\and

\author{Julian H. Krolik}
\affil{Department of Physics and Astronomy, Johns Hopkins University, 
    Baltimore, MD 21218}

\begin{abstract}

We have employed a 3-d energy-conserving radiation MHD code
to simulate the vertical structure and thermodynamics of a shearing
box whose parameters were chosen so that the radiation and gas pressures
would be comparable.  The upper layers of this disk segment are
magnetically-dominated, creating conditions appropriate for both
photon bubble and Parker instabilities.  We find little evidence for
photon bubbles, even though the simulation has enough spatial resolution
to see them and their predicted growth rates are high.   On the other hand,
there is strong evidence for Parker instabilities, and they appear to
dominate the evolution of the magnetically supported surface layers.
The disk photosphere is complex, with large density inhomogeneities
at both the scattering and effective (thermalization) photospheres of the
evolving horizontally-averaged structure.  Both the dominant
magnetic support and the inhomogeneities are likely to have strong
effects on the spectrum and polarization of thermal photons emerging
from the disk atmosphere.  The inhomogeneities are also large enough
to affect models of reflection spectra from the atmospheres of accretion
disks.

\end{abstract}


\section{Introduction}

    The first attempts to understand accretion disks concentrated on deriving
radial profiles of azimuthally-averaged and vertically-integrated quantities
in time-steady disks.  For example, \citet{sha73} and \citet{nov73} worked out the
theory for such disks around black holes by applying the constraints of energy
and angular momentum conservation.
When \citet{sha73} introduced the notion that the
stress responsible for moving matter inward through a disk should be
proportional to the pressure, it was with a view toward tying
the stress to local conditions in an azimuthally-averaged and
vertically-integrated sense.  Efforts were made within that framework
to compute the actual vertical thickness as a function of radius, but it
was recognized that this computation was subject to considerable uncertainty
due to lack of knowledge of internal disk dynamics.

Particularly with the recognition that disk stresses
can largely be explained by correlated MHD turbulence stirred by the
magneto-rotational instability (as reviewed, for example, in \citealt{bal98}),
recently attention has been given to what may happen inside
disks, even when their vertical scale-heights are quite small compared
to the distance to the central object.  This attention has included both
analytic work on wave modes and instabilities that may exist (e.g.
\citealt{now93,gam98,kim00,pes05}) and numerical simulations of vertically
stratified shearing-box segments of disks \citep{bra95,sto96,mil00,tur04,hir06}.

       Special problems arise in understanding the vertical structure of
disks when radiation pressure is comparable to or greater than gas pressure.
As first pointed out by \citet{sha73}, this condition is likely
to occur in the inner portions of disks accreting at more than a small fraction
of Eddington, particularly when the central mass is $\gg 1 M_{\odot}$.
Perhaps the biggest potential problem is that thermal and
viscous instabilities in the vertically-integrated quantities might
occur as a result of radiation pressure dominance \citep{lig74,sha76}.

The large radiation pressure regime may also be vulnerable to smaller scale
photon bubble instabilities
\citep{gam98}.  The most rapidly-growing form of this instability occurs
at short wavelengths, where photons diffuse rapidly, resulting in
radiative amplification of magnetosonic waves \citep{bla03}.
At least for strong magnetic fields, the initial nonlinear evolution of the
instability in this short wavelength limit produces trains of shocks
\citep{beg01,tur05}.  \citet{beg02,beg06} has suggested that these shock
trains would permit accreting black holes to radiate at luminosities 1-2
orders of magnitude larger than Eddington.

       We are embarked on a program to explore the nature of the
internal dynamics in disks at varying levels of relative importance of
radiation pressure support.  We do this by conducting detailed numerical
simulations that both conserve energy to
high accuracy and include radiation transport in a vertically stratified
disk annulus.  The first such simulation was performed by \citet{tur04}
with a high ratio of radiation to gas pressure at the midplane.  While very
interesting results were obtained, the
simulation achieved only partial energy conservation and was not able to
reach low enough densities to incorporate the photosphere within the simulation
domain.  Solutions to these problems have since been found and a simulation
of a disk annulus in which the radiation pressure was about $20\%$ of the gas
pressure was presented by \citet{hir06}.  In a companion paper to this one
\citep{kro07}, we report on the
thermodynamic properties of a simulated disk annulus in which the time-averaged
radiation and gas pressures were roughly equal.

       In this paper, we investigate internal structures near the
surface of the same disk annulus whose thermodynamics were studied in
the companion paper.  We concentrate on these upper
regions because the shorter length scale instabilities predicted by linear
theory tend to have their fastest growth rates there.   As we shall show,
two are of special interest: the photon bubble instability and the Parker
instability.  In addition, it is the structure of these surface layers that
is most important in determining the degree to which the photon spectrum
emerging from the disk deviates from a black body at the local effective
temperature.

This paper is organized as follows.  In section 2 we briefly summarize the
methods and parameters of the simulation we are investigating.  Then in
section 3 we describe the properties of photon bubble and Parker instabilities
that we expect based on applying linear instability theory to the horizontally
averaged structure in the simulation.  In section 4 we compare these
expectations to what we actually observe in the three-dimensional
structure of the simulation.  Strong inhomogeneities are observed near
the photosphere in the simulation, both in density and, to a lesser extent,
temperature, and we describe these in section 5.  In section 6, we discuss
the possible implications of our results for models of the emergent and
reflected photon spectra, and for the emergent photon polarization.  Finally,
we summarize our results in section 7.

\section{Calculational Method}

     Our results are based on a 3-d radiation MHD simulation of a vertically
stratified section of an accretion disk, computed under the shearing-box
approximation.  The code is described in greater detail in \citet{hir06},
although the version we used differs slightly from that one by certain
technical improvements.  In its essential elements, this code computes
the equations of 3-d ideal MHD, capturing all grid-scale numerical dissipation
as heat.  Radiation transport and radiation forces are included, but with
thermally-averaged opacity and emissivity and in the flux-limited diffusion
approximation.  Further details about initial conditions and boundary conditions
are given in the companion paper \citep{kro07}, which concentrates on
describing the thermodynamics and larger-scale properties of this simulation.

In order to arrive at a disk annulus in which the radiation and gas pressures
are comparable, we chose the following parameters for this shearing-box segment:
The central mass is $6.62 M_{\odot}$, the radius is
$r=150r_g = 1.5 \times 10^8$~cm,
and the surface density is $4.7 \times 10^4$~g~cm$^{-2}$.  If it were described
by a Shakura-Sunyaev stress ratio $\alpha = 0.02$, the mass accretion
rate would produce a luminosity $0.1L_E$ if the efficiency were 0.1.
The local effective temperature would then be $9.0 \times 10^5$~K.
Our unit of length $H=3.1 \times 10^6$~cm is the (half)-disk thickness
in the gas-dominated limit, as given in \citet{kro99}.
The midplane optical depth
is $(1.58 \pm 0.01)\times 10^4$, varying slightly due to the
temperature-dependence of the free-free contribution.  At all times,
the opacity is dominated by Thomson scattering.  The initial magnetic
field is a twisted azimuthal flux tube with no net poloidal flux and
energy density $1/25$ the midplane gas plus radiation pressure.

Our computational box had a vertical thickness of $12H$, stretching
symmetrically above and below the midplane.  Its radial thickness was
$(3/4)H$ and its azimuthal thickness was $3H$.   These dimensions were
described by $512 \times 32 \times 64$ cells, respectively.  The simulation
ran for 318 orbits, equivalent to more than 40 thermal times.  As discussed
in the companion paper \citep{kro07}, the epoch with highest total energy
content was at 90 orbits, and the epoch with lowest total energy content
was at 150 orbits.  We focus our analysis in this paper on just these two
epochs.

\section{Expectations Based on Linear Instability Theory}

\subsection{Photon bubbles}

To check for the presence of photon bubble instabilities in the present
simulation, we must first delineate the regions where photons diffuse
rapidly relative to the time scales relevant
for acoustic waves in a gas and radiation mixture \citep{ago98}.
When gas and radiation exchange heat rapidly, an excellent approximation
for most, but not all, regimes of plasma acoustic behavior in the case
under consideration here, it is convenient to write the dispersion relation
for compressive waves in terms of
an effective sound speed (e.g.  Appendix B of \citealt{bla03})
\begin{equation}
C_s^2={\omega\left[e+4E\right]c_{\rm t}^2+
{ick^2\over3\kappa_{\rm F}\rho}4Ec_{\rm i}^2\over
\omega\left[e+4E\right]+{ick^2\over3\kappa_{\rm F}\rho}4E}.
\label{eq:csdiff}
\end{equation}
Here $k=2\pi/\lambda$ is the wavenumber of the disturbance, $\omega=kC_s$
is the corresponding angular frequency, $e$ is the internal energy density
of the gas, $E$ is the radiation energy density, $\rho$ is the density,
and $\kappa_{\rm F}$ is the flux mean opacity.  The quantity
$c_{\rm i}=(p_{\rm g}/\rho)^{1/2}$ is the isothermal sound speed in the gas,
where $p_{\rm g}$ is the gas pressure.  The total adiabatic sound speed
$c_{\rm t}$ of the fluid is defined by
\begin{equation}
c_{\rm t}^2={16E^2+60(\gamma-1)Ee+9\gamma(\gamma-1)e^2\over9(e+4E)\rho}
={\Gamma_1(p_{\rm g}+E/3)\over\rho},
\end{equation}
where $\gamma=5/3$ is the adiabatic index of the gas and $\Gamma_1$ is the
first generalized adiabatic exponent relating pressure and volume changes
when radiation and matter are tightly thermally coupled \citep{cha67}.
Equation (\ref{eq:csdiff}) shows that
the wave speed ranges from $c_{\rm i}$ to $c_{\rm t}$; the frequency for a given
wavelength may therefore be easily bounded.

It is then straightforward to show
that acoustic disturbances with wavelengths satisfying
\begin{equation}
\lambda<{2\pi c\over 3\kappa_{\rm F}\rho c_{\rm i}}\left({4E\over e+4E}\right)
\left({c_{\rm i}\over c_{\rm t}}\right)^2\equiv\lambda_{\rm R}
\label{eq:lamr}
\end{equation}
are in a regime where photons diffuse rapidly compared to the wave period.
As a result, the radiation pressure response is lost, and the sound speed
approaches the isothermal sound speed in the gas $c_{\rm i}$.
On the other hand, acoustic disturbances with longer wavelengths satisfying
\begin{equation}
\lambda> \lambda_{\rm S} \equiv
\frac{c_{\rm t}}{c_{\rm i}} \lambda_R 
= \left[\Gamma_1\left(1 + \frac{E}{3p_{\rm g}}\right)\right]^{1/2}\lambda_R
\end{equation}
trap photons, and therefore propagate at $c_{\rm t}$,
the total sound speed in the gas and radiation.  For intermediate wavelengths
with $\lambda_{\rm R}<\lambda<\lambda_{\rm S}$, the damping of the
disturbances is maximal.  However, when the radiation energy density is not
much larger than the gas pressure, as in the present simulation, the maximal
damping rate is less than the mode period, and $c_{\rm t}\sim c_{\rm i}$.

\begin{figure}
\epsscale{1.0}
\plotone{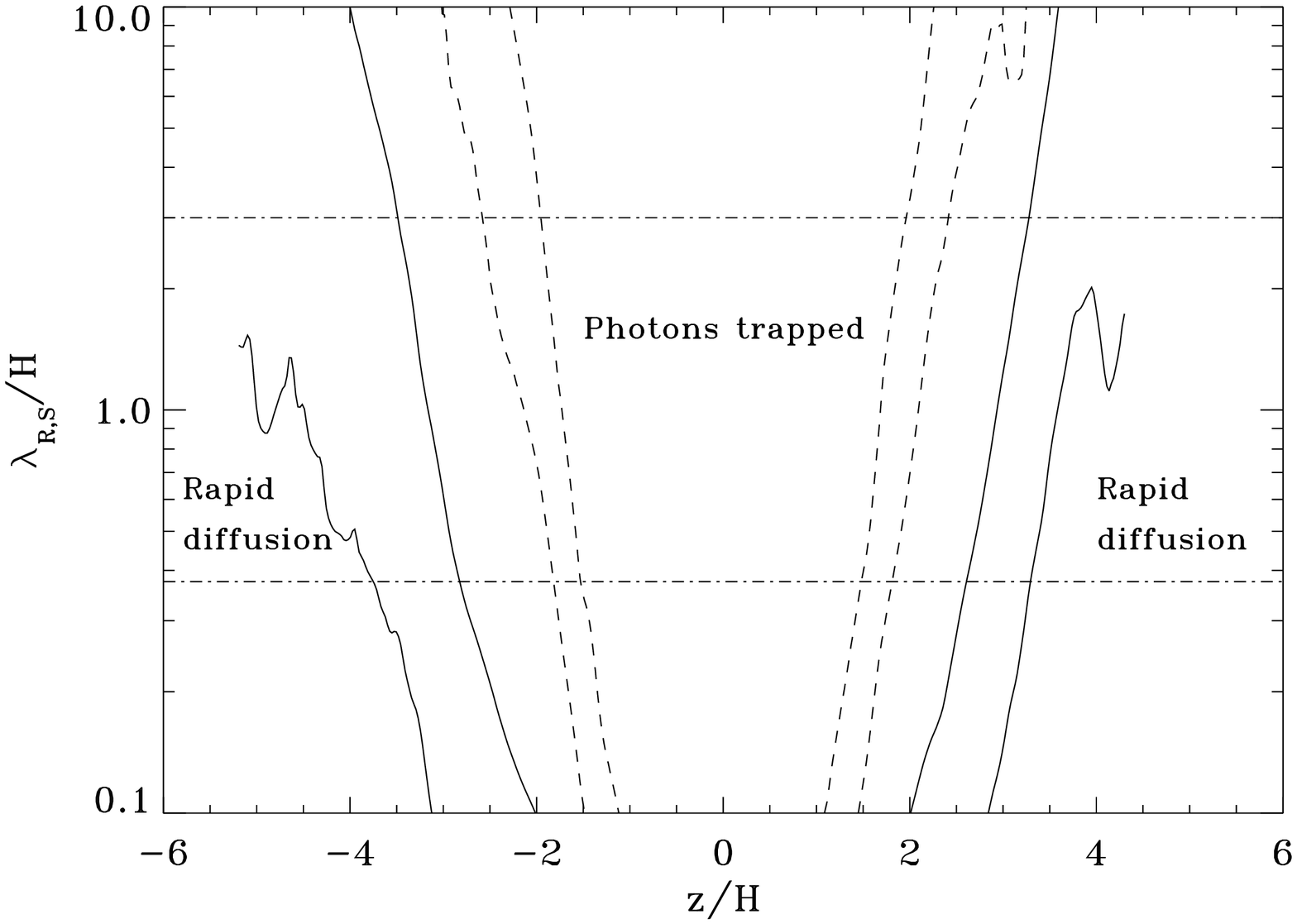}
\caption{Wavelength $\lambda_{\rm R}$ below which rapid photon diffusion
applies (lower of each pair of curves), and wavelength $\lambda_{\rm S}$ above
which photons are trapped (upper of each pair of curves) for acoustic
disturbances in the simulation.
The solid and dashed pair of curves correspond to horizontally averaged
conditions at the highest ($t=90$ orbits) and lowest ($t=150$ orbits) total
energy epochs, respectively. The curves extend only out to the Rosseland
mean photosphere of the horizontally averaged structure
in each case, outside of which photon diffusion does not apply.  For
comparison, the lower horizontal dot-dashed line shows the length
$8r\Delta\phi$ of eight grid zones in the azimuthal direction, close to
the minimum wavelength that the code can resolve.  The upper horizontal
dot-dashed line shows the length $3H$ of the box in the azimuthal
direction.
\label{fig:rlamrsplot}}
\end{figure}

Fig.~\ref{fig:rlamrsplot} shows the critical wavelengths $\lambda_{\rm R}$
and $\lambda_{\rm S}$ as a function of height at the highest (90 orbits)
and lowest (150 orbits) total energy epochs in the simulation, based on
horizontally averaged data.  Also indicated is the range of wavelengths
in the azimuthal direction that can be resolved by
the code, from eight azimuthal cell sizes to the box size in the azimuthal
direction.
Near the midplane, photons are trapped at all epochs for acoustic disturbances
at all wavelengths resolved by the code.
Outside $\pm2.5H$, compressive disturbances
are in the rapid diffusion limit for all accessible wavelengths at $t=150$
orbits, when the disk is relatively cool.  Even at $t=90$ orbits, when the
disk is relatively hot, there is a factor of two range of resolved wavelengths
that are in the rapid diffusion limit in the subphotosphere surface layers.

\begin{figure}
\epsscale{1.0}
\plotone{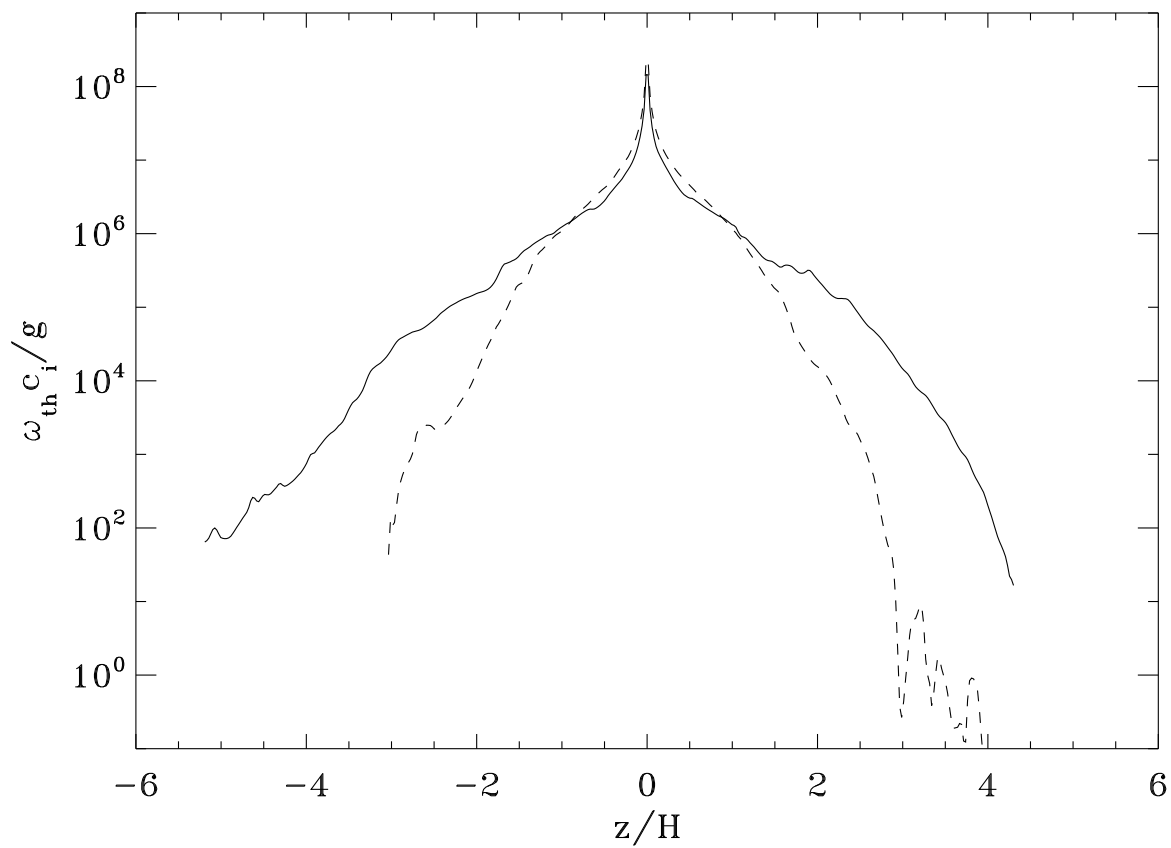}
\caption{Scaled photon/gas thermal coupling frequency calculated from the
horizontally averaged structure at $t=90$ orbits (solid) and $t=150$ orbits
(dashed).  Tight thermal coupling is a good approximation for photon bubbles
if this scaled frequency is much greater than unity.}
\label{fig:pbiomthplot}
\end{figure}

In addition to the rapidity of photon diffusion, another issue of relevance
to photon bubbles is how rapidly gas and radiation exchange heat.  As
discussed in \citet{bla03}, the gas and radiation temperatures are effectively
locked together in a rapidly growing photon bubble disturbance
provided $\omega_{\rm th}\equiv4E\kappa\rho c/e$ is greater than
the gas acoustic frequency $g/c_{\rm i}$ associated with a gas pressure scale
height.  (Here $\kappa$ is the Planck mean opacity associated with
the free-free absorption used in the code.)  The ratio of these two rates is
shown in Fig.~\ref{fig:pbiomthplot}, and except for the
uppermost layers at 150 orbits, rapid heat exchange is a good approximation
for photon bubbles.

\begin{figure}
\epsscale{1.0}
\plotone{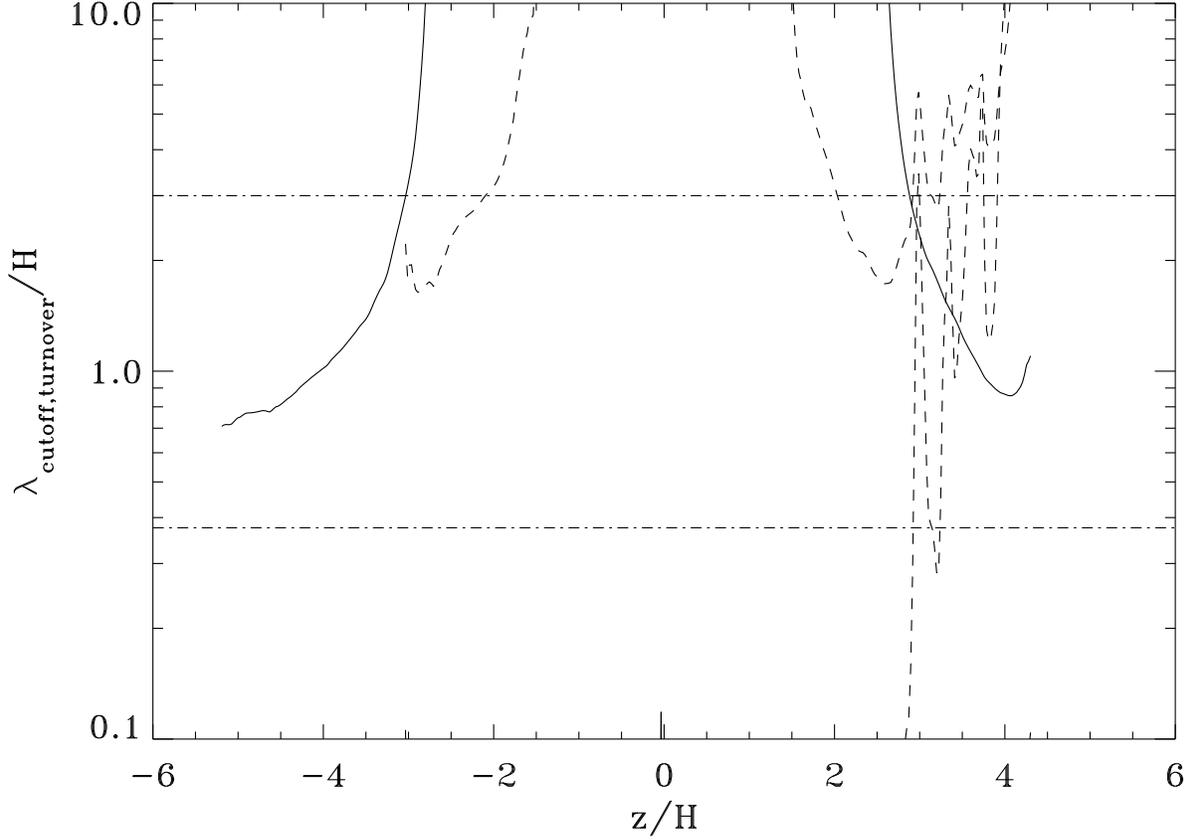}
\caption{Range of wavelengths over which the asymptotic photon bubble
growth rates of Fig.~\ref{fig:pbigrowthrate} should be valid (provided these
wavelengths are also in the rapid photon diffusion regime), as a function of
height in the box at $t=90$ orbits (solid) and $t=150$ orbits (dashed).
The solid and upper dashed curves in each epoch correspond to the turnover
wavelength, while the lower dashed curve on the right corresponds to the
thermal cutoff wavelength at $t=150$ orbits.  (The thermal cutoff
wavelength for $t=90$ orbits is less than $0.1H$ at all heights, and is
also less than $0.1H$ for $z<2.8H$ at $t=150$ orbits.)
Rapid growth rates should exist between the thermal cutoff and turnover
wavelengths.
The lower and upper horizontal dashed lines again indicate eight
cell sizes $8r\Delta\phi$ and the length $3H$ of the box in the azimuthal
direction, respectively.}
\label{fig:rlamturn}
\end{figure}

In this thermally locked, rapid diffusion regime, photon bubbles reach their
maximum growth rates $\gamma_d$ over a finite range of wavelengths.  The
approximate lower bound is a short thermal ``cutoff wavelength"
$\lambda_{\rm cutoff}=2\pi c_{\rm i}/(\omega_{\rm th}\gamma_d)^{1/2}$
below which rapid heat exchange
fails to be a good approximation in the amplifying acoustic wave, and the wave
stabilizes due to the damping associated with the slow heat exchange.
The approximate upper bound is a long ``turnover wavelength"
$\lambda_{\rm turnover}=\gamma_d/v_{\rm ph}$
above which the wave is no longer a simple magnetosonic wave propagating
with phase speed $v_{\rm ph}$, and the growth rate declines.
These wavelength bounds are shown in Fig.~\ref{fig:rlamturn} for the
horizontally-averaged conditions in the simulation.

\begin{figure}
\epsscale{1.0}
\plotone{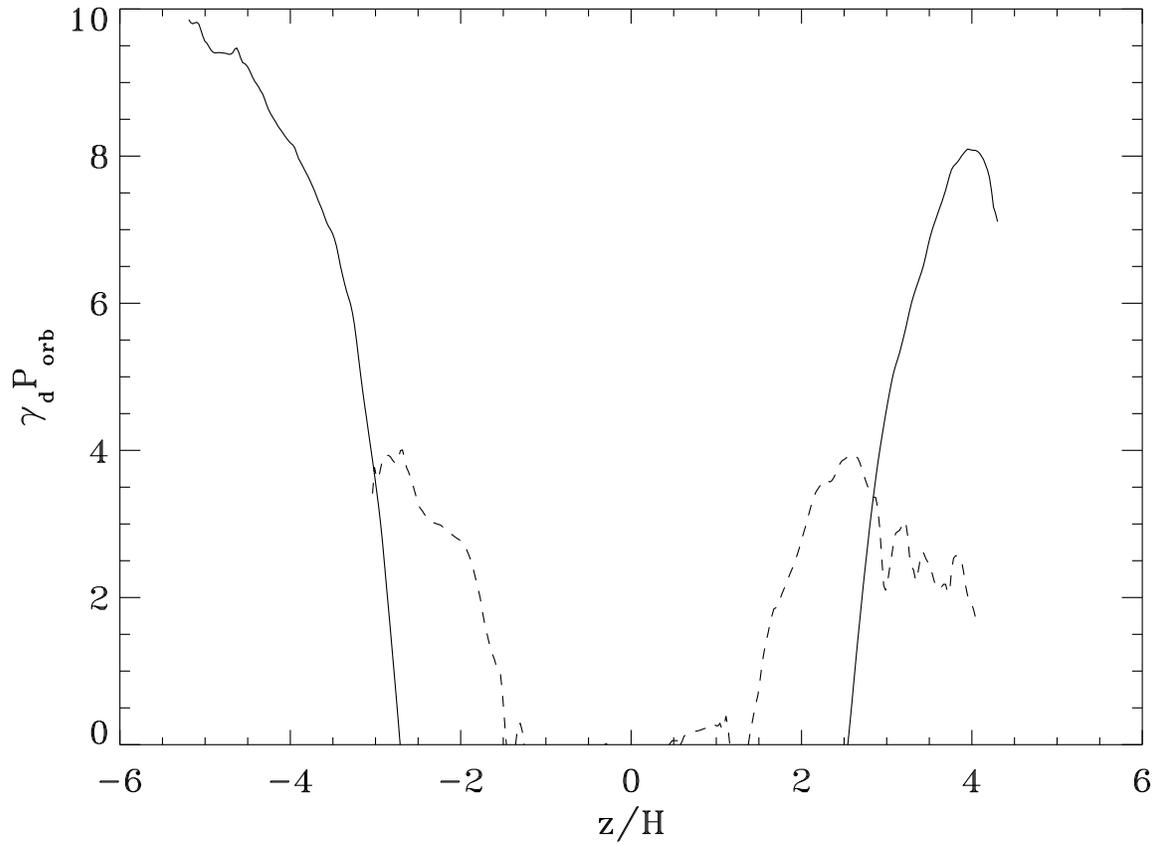}
\caption{Short wavelength photon bubble growth rate $\gamma_d$ scaled with the
orbital period $P_{\rm orb}=2\pi/\Omega$, as a function of height at $t=90$
orbits (solid) and $t=150$ orbits (dashed).
Growth rates are shown only at heights
deeper than the photosphere of the horizontally averaged
structure at that epoch.
\label{fig:pbigrowthrate}}
\end{figure}

Fig.~\ref{fig:pbigrowthrate} shows the expected maximum growth rate $\gamma_d$
of the photon bubble instability (Eq. 93 of \citealt{bla03})
as a function of height in the simulation box at 90 and 150 orbits.
We assumed that the relevant wavelengths are in the rapid diffusion limit,
which Figs.~\ref{fig:rlamrsplot} and \ref{fig:rlamturn} show to be valid
at least in the subphotospheric surface layers.
We also assumed a purely azimuthal magnetic field and a wave vector
orientation inclined upward at 45 degrees to the horizontal in the
vertical/azimuthal plane.
Much of the field is in fact in the azimuthal direction in the upper layers
and 45 degrees is close to the inclination that gives maximal growth.
Apart from particular cases where the growth rate strictly vanishes, other
field orientations and wave vector directions alter the growth rates by
geometric factors of order unity.  It is important to note that we
calculate growth rates only inside the Rosseland mean photosphere of the
horizontally averaged structure, as the photon bubble instability
does not exist when the medium is optically thin.

Figs. 3 and 4 show that at $t=90$ orbits, modes with wavelengths shorter
than about a scale height should grow by many $e$-foldings per orbital period
in the surface layers beneath the Rosseland mean photosphere.
Our simulation has plenty of resolution to see
these rapid growth rates:  a scale height projected through $45^\circ$
corresponds to 15 zones in the $\phi$-direction and 30 zones in the
$z$-direction.  Resolution studies by \cite{tur05} found that the growth
rates predicted by linear theory should be recovered in a numerical simulation
to within ten percent provided there are more than 15 zones per wavelength.
On the other hand, Fig. 3 also shows that wavelengths more than a factor of
two smaller than
the turnover wavelength are not well-resolved.  Figs.
6 of \citet{bla03} and 12 and 13 of \citet{tur05} show that the growth
rate at the turnover wavelength can be smaller by as much as a factor of
two compared to the maximum growth rate $\gamma_d$, but that still gives
$4-5$ $e$-foldings per orbital period.
The maximum growth rates are considerably smaller at $t=150$ orbits,
and will be reduced further above $z=2.5H$ in the upper half of the simulation
due to the breakdown in good thermal coupling between the gas and radiation.
Photon bubbles are therefore most likely to be present at $t=90$ orbits.

\subsection{The Parker instability}

All analyses of the magnetosonic photon bubble instability so far have assumed
a uniform magnetic field in the equilibrium \citep{gam98,beg01,bla03}.  Even
if that field is strong, it does not support the background against gravity.
In contrast, we have shown in section~\S 3.2 of the companion paper
\citep{kro07} that the magnetic
field in our simulation provides the dominant support against gravity at high
altitudes.  This fact immediately suggests that the Parker
instability \citep{par66,par67} might play a role instead of or in addition
to the photon bubble instability.

We can get a rough idea of the relevance of
the Parker instability to this simulation by examining its dispersion relation
in an isothermal atmosphere.  We do this in the Appendix.
Figs.~\ref{fig:parkergrowthrate} and \ref{fig:rlammax}
show the maximum growth rate [equation (\ref{eq:gammamax})] and characteristic
wavelengths [equations (\ref{eq:lamparker})-(\ref{eq:lammax})] of the Parker
instability as a function of height at $t=90$ and 150 orbits.
These figures are based on horizontally averaged data at each epoch, and on
the assumption that the
magnetic field is azimuthal.  We also assumed that radiative diffusion is rapid,
so that perturbations in the gas are isothermal, though we believe that the
results are approximately valid even under conditions where the radiation
is trapped.  This is because the Parker instability arises from a competition
between magnetic pressure and thermal pressure, and the gas and radiation
pressures are comparable here.  Fig.~\ref{fig:rlammax} leads us to expect
that the regions outside approximately $\pm2H$ will be Parker unstable during
the high total energy epoch at $t=90$ orbits.  In the low total energy
epoch at $t=150$ orbits, we expect the regions $z<-2H$ and $H<z<3H$ to be
Parker unstable.

The Parker instability growth rates exceed the orbital frequency by roughly
a factor of two, and are larger than the maximal photon bubble
growth rates except perhaps very near the Rosseland mean photosphere
at $t=90$ orbits.
Interestingly, the wavelength for maximal growth of the Parker
instability roughly corresponds to the azimuthal size of the box, and
our analysis in the Appendix indicates that the corresponding wave vector
should be in the horizontal direction.

\section{To what extent are instabilities present in the simulation?}

\subsection{Photon bubbles}

The simulation results are somewhat ambiguous as to the presence of
the photon bubble instability.
Fig.~\ref{fig:rhoyzupper090} shows the density, magnetic field, and velocity
field in the vicinity of the upper photosphere at 90 orbits, the time of
highest total energy in the simulation.  Two photospheres are actually
indicated in this figure, both calculated from the horizontally averaged
structure at this epoch:  the Rosseland mean photosphere and the effective, or
thermalization photosphere.  The latter is the outermost surface where photons
exchange energy with matter.  It is therefore the surface inside of which
photons can come into thermal equilibrium with the matter.  Because Thomson
scattering dominates the free-free absorption opacity throughout the domain
of the simulation, the effective photosphere is significantly deeper than
the Rosseland mean photosphere.

In the region $3H<z<4H$ there are 2-3 relatively
high density regions which are somewhat reminiscent of the shock trains
that form in the initial nonlinear development of the photon bubble
instability \citep{beg01,tur05}.  The high density streak stretching
diagonally from the upper
left ($-0.5H$,$3.8H$) to the lower right ($0.7H$,$3.2H$) in the center of the
figure is almost certainly a shock, as the velocity field has significant
convergence in this region.
Two other shock-like features in the lower left
of the figure [ranging from ($-0.9H$,$3.2H$) to ($-0.5H$,$3.0H$) and from
($-1.1H$,$3.1H$) to ($-0.7H$,$2.8H$)] run parallel to the first shock, although
whether these are true
shocks is not completely clear as the velocity field does not show as much
convergence as at the first shock.
The magnetic field lines show a fairly discontinuous change in orientation
across the first shock, which might be indicative of some buckling under the
weight of the dense material.  There is also a
rapid change in orientation of the magnetic field lines near the other two
putative shocks.
Examination of nearby radial
slices at the same time shows that the first shock extends over
only $\simeq 0.2H$ in radius.  Our simulation data dumps lack the
time resolution to explore its evolution: it is absent and the overall
structure is quite different one orbit earlier and one orbit later.
This is not surprising, however, as the background differential rotation will
shear out a region of radial thickness $x=0.2H$ to extend over
$3\pi x\simeq2H$ in the azimuthal direction after one orbit.

It is unclear whether these shocks are the result of photon bubble instability.
To try to resolve this ambiguity, we have investigated how the
radiation flux and fluid velocity are correlated in the layers where the
photon bubble instability is expected.  At least in the rapid diffusion limit
of interest here, photon bubbles are fundamentally a radiative amplification
of a magnetosonic wave.  The density compressions and rarefactions in the wave
alter the diffusion paths of photons in such a way as to produce radiation
flux perturbations that everywhere make acute angles with the fluid velocities
in the wave \citep{tur05}.  Oscillating radiation pressure
forces in phase with the oscillating fluid velocities are the result, which
drive the wave to larger amplitudes.  To check for this distinctive
directional relationship between the perturbed forces and velocities, we
show in Fig. \ref{fig:pbidiagnostic090}
the distribution of cosines of the angle $\theta$ between the fluctuating
radiation flux and fluid velocity vectors at all grid points in a set of
horizontal slices through the simulation at 90 orbits.  To be precise,
\begin{equation}
\cos\theta={({\bf F}-<F_z>\hat{\bf z})\cdot({\bf v}+{3\over2}\Omega
x\hat{\bf y})\over|{\bf F}-<F_z>\hat{\bf z}||{\bf v}+{3\over2}\Omega
x\hat{\bf y}|},
\end{equation}
where the angle brackets refer to a horizontal average at the particular height
in question.  We subtract off the horizontal average of the vertical flux
component because a radiation driven outflow would also give rise to a
correlation between flux and velocity without having
anything to do with photon bubbles.

As can be seen in Fig.~\ref{fig:pbidiagnostic090}, at three of the
four sample heights, there is essentially no directional correlation
between the fluctuating part of the radiation flux and the fluid velocity.
Only at $z=+2.4H$ is there a significant positive correlation; in fact,
there is a weak anti-correlation at $z=+3.3H$ and $3.8H$.  Ironically,
our application of the linear theory of photon bubbles to these data
suggests that they grow only for $z > 2.5H$.  We conclude,
therefore, that there is little evidence for photon bubble-driven
dynamics here.

\subsection{Parker instability}

Although evidence for photon bubbles is scant, evidence for Parker
instability is strong.  This mode involves an interplay between
support against gravity provided by magnetic pressure forces and
restraint against magnetic buoyancy provided by magnetic tension
forces.  We therefore begin this part of our analysis by mapping
out the strengths of these forces.

The vertical accelerations produced by magnetic
tension and magnetic pressure are
\begin{equation}
a_T={1\over4\pi\rho}\left(B_x{\partial B_z\over\partial x}+
    B_y{\partial B_z\over\partial y}\right)
\end{equation}
and
\begin{equation}
a_P={-1\over8\pi\rho}{\partial\over\partial z}\left(B_x^2+B_y^2\right),
\end{equation}
respectively.  These accelerations are plotted in Fig.~\ref{fig:vertaccmag}.
In a volume-averaged sense, the magnetic pressure force almost always pushes
vertically outward; by contrast, in the same sense of averaging,
the magnetic tension force generally holds matter in.  In this volume-averaged
sense, the two magnetic forces appear almost to balance each other.  As we
shall see shortly, in Parker modes there is a strong correlation between the
sense of curvature of the magnetic field (which controls the direction of the
tension force) and the gas density: relatively low density and inward
tension force are often found together.  For fixed total mass, low density
regions must always occupy more total volume than those of high density,
so inward magnetic tension regions dominate in a volume average.  On
the other hand, support against gravity is best thought of in terms of
mass-weighted quantities (right-hand panel of Fig.~\ref{fig:vertaccmag}).
Averaged in this fashion, magnetic tension becomes much less important,
and we see that the upper layers of the disk are, as previously noted,
primarily supported by magnetic pressure gradients.

The fact that large inward magnetic tension forces begin outside $z=\pm3H$
is an indication that the Parker instability is in operation at those
altitudes.  This height
is not far from the $z=\pm2H$ boundaries where we expected
the Parker instability to develop on the basis of our simplified linear
analysis.


A further diagnostic of Parker instability can be seen in the density and
magnetic field structure shown in Fig.~\ref{fig:rhoyzupper090} at $t=90$
orbits.  The densest regions of
gas lie in troughs of magnetic field lines, whereas the lighter regions are
located near magnetic field line crests.  To quantitatively check this visual
impression, the left hand panel of Fig.~\ref{fig:tensionvsrho} shows the
magnetic tension force per unit volume versus the
density at each grid zone in a $z=2.75H$ horizontal slice through the box
at this epoch.
There is a clear correlation between these two quantities: gas
heavier (lighter) than average is predominantly associated with upward
(downward) tension forces.

Fig.~\ref{fig:rhoyzupper150} shows the upper surface layers at $t=150$
orbits.  The structure above the photosphere is very complex with large
flow velocities, and the $\phi-z$ projection of the field shows circulation,
including regions of nearly vertical field lines.  This complex structure
is no doubt related to the large deviations from hydrostatic balance
(see right panel of Fig.~6 in the companion paper, \citealt{kro07}).
Nevertheless,
a long wavelength Parker mode appears to be present in the region $2H<z<3H$,
which is again where it was expected to occur based on the minimum unstable
wavelength shown in Fig.~\ref{fig:rlammax}.  Its wavelength and horizontal
orientation also conform with the linear analysis expectations.
The right hand panel of Fig.~\ref{fig:tensionvsrho} shows that tension is
again highly correlated with density in the $z=2H$ horizontal slice.

This correlation between density and tension forces is perhaps the strongest
evidence that Parker modes play a significant role in the outer layers
of the simulation.  A volumetric rendering of the correlation is shown in
Fig. ~\ref{fig:fieldlines}.  Regions that are denser than average at a given
height are mostly associated with field line valleys, whereas low density
regions are mostly associated with crests in the field lines.

Fig.~\ref{fig:tensionrhostats} shows the fractional horizontal surface
area at each height for which the correlation between density and tension
forces holds.  Within approximately two scale heights on either side of
the midplane, there is no correlation.  Outward and inward tension forces
in this region are equally likely to be associated with higher and lower
than average densities, so the fractional surface area exhibiting the
expected behavior of the Parker instability is fifty percent.  Further
away from the midplane at $t=90$ orbits, however, the vertically outward
(inward) tension
forces are more likely to be associated with lower (higher) than average
density, as expected from the Parker instability.  This vertical dependence
of the magnetic tension-density correlation is completely consistent with
our linear instability expectations from section 3.2 above: unstable
growth at $|z| \gtrsim 2H$ and no instability closer to the midplane.
The situation is approximately the same at $t=150$ orbits, except that there
is no evidence of the expected Parker correlation below $z\simeq-3H$, or
near $z=3H$, $4H$, and above $5H$.  Our linear analysis of section 3.2
predicted that at this epoch the regions $z<-2H$ and $H<z<3H$ would be
Parker unstable.
Again, we suspect that the discrepancy arises partly from the breakdown of
hydrostatic equilibrium in the outer layers at this epoch, as shown in
the right panel of Fig. 6 of the companion paper.

It is fortunate that our simulation domain just happened to be the right
size for the maximal Parker instability wavelengths to fit inside the box.
Whether these modes can fit in the computational domain is clearly something
to be borne in mind when choosing parameters for future stratified shearing
box simulations.  It is unclear whether the finite azimuthal extent of the
box might be affecting the nonlinear saturated state of the instability.  One
could imagine, for example, that the tension forces
could be reduced if the nonlinear development of the
instability cascaded toward longer wavelengths, but this would not be
possible in our finite shearing box.  Clearly further work with larger
simulation domains is warranted in the future.

For comparison, the upper surface layers in the middle of the box near the
end of the gas pressure dominated simulation of \citet{hir06} are shown in
Fig.~\ref{fig:rhoyzupperpgas}.  The structure in the lower regions of this
figure again appears to be consistent with the Parker instability.
Fig.~\ref{fig:pgastensionrhostats} shows that the expected correlation
between density and tension forces is again widespread outside approximately
two scale heights away from the midplane.  Not surprisingly, the Parker
instability appears to be generically important in the
magnetically supported upper layers of accretion disks even when there is
little radiation pressure.

\section{Photospheric Irregularity}

Very large, irregular density inhomogeneities
are evident in the surface layers and extend as deep as the
effective photosphere of the horizontally-averaged structure at all epochs.
The distribution of densities at the upper effective and Rosseland mean
photospheres at 90 and 150 orbits is shown in Fig.~\ref{fig:rhofluct}.
The density fluctuations range over factors of approximately ten above and
below the mean, with small isolated rarefied regions that are even less
dense still.
The vertical distribution of horizontally-averaged root mean square fractional
fluctuation of density is shown in Fig.~\ref{fig:rhormshave} for the same
epochs.  The large density inhomogeneities clearly extend over a broad range
of heights outside $\pm2H$.

One possible explanation for the presence of these density inhomogeneities
lies in the fact that, in isolated parts
of the most rarefied regions outside the effective photosphere, the gas
temperature can greatly exceed (by greater than a factor 100) the effective
temperature of the radiation, as shown in Figs. \ref{fig:tgasotrad090} and
\ref{fig:tgasotrad150}.  The code assumes that gas and radiation exchange
heat purely through free-free absorption and emission, and the density
is so low in these regions that the gas and radiation are unable to
equilibrate through this channel.  Including Compton scattering in the
gas and radiation energy equations would enhance the thermal coupling.
The ratio of time scales of Compton to free-free cooling of the gas
is approximately
\begin{equation}
{t_{\rm Comp}\over t_{ff}}\simeq{\kappa(T_{\rm gas}^4-T_{\rm rad}^4)
m_ec^2\over4k_{\rm B}(T_{\rm gas}-T_{\rm rad})\kappa_TT_{\rm rad}^4},
\end{equation}
where $\kappa$ is the Planck mean free-free opacity, $\kappa_T$ is the
Thomson opacity, $T_{\rm gas}$ is the gas temperature, and $T_{\rm rad}$
is the effective temperature of the radiation.  In the conditions of the
current simulation, free-free is faster than Compton in exchanging heat
between the gas and radiation throughout most of the volume.  However,
in the isolated regions where the gas temperature significantly exceeds the
radiation temperature, the Compton cooling time can be shorter than the
free-free cooling time by factors of 5 to 10.  
The gas may therefore be artificially too hot in these regions, because
we did not include Compton cooling in the simulation.  Artificially
elevated temperature might enhance the degree of inhomogeneity by
artificially raising the gas pressure.  Although this effect may explain
the regions we see where the gas and radiation temperatures are far
apart, it cannot explain all regions of large density inhomogeneity:
Figs. \ref{fig:tgasotrad090} and \ref{fig:tgasotrad150} show that the
radiation and gas temperatures are equal at the effective photosphere
(as they must be), and yet there can be large density contrasts there.

Photon bubbles would produce inhomogeneities, but we have failed to find
any other evidence for these instabilities in the simulation.  Moreover,
substantial inhomogeneities persist even in the optically thin uppermost
regions, where the photon bubble instability cannot act.
Figs.~\ref{fig:rhofluct} and \ref{fig:rhormshave} show that strong density
inhomogeneities are also present in the gas pressure dominated simulation
of \citet{hir06}.  Photon bubbles are almost certainly
irrelevant in this case, and therefore cannot be
responsible for the inhomogeneities that are seen.

As noted by \citet{hir06}, a more likely explanation for the inhomogeneities
is that they are a result of the fact that magnetic forces dominate gas
and radiation pressure in these layers.  Because density only affects gas
pressure gradients, and these generally play only a small role in the dynamics
in these regions, it is not surprising that density inhomogeneities develop
in response to fluctuating magnetic forces.

\section{Implications for Emergent and Reflected Photons}

        Our earlier result that magnetic support extends the upper atmospheres
of accretion disks when gas pressure dominates \citep{hir06} now appears
to apply to conditions in which radiation pressure is comparable to gas
pressure.
This additional source of support against gravity creates a lower density
photosphere than would otherwise be found.  As a result, electron scattering is
enhanced over absorption, and the matter's ionization balance shifts toward
more highly-ionized species, weakening absorption edges and strengthening
emission edges.  Both of these effects generically lead to a {\it hardening}
of the emergent spectrum \citep{bla06}.  On the basis of the calculations
reported here, we expect that magnetically supported, extended atmospheres
should continue to exist in rings even where radiation and gas pressure are
comparable.

However, we also see strong, irregular density inhomogeneities
in the surface layers beneath the scattering photosphere, and these
inhomogeneities were also present in the gas pressure dominated simulation
(Figs.~\ref{fig:rhoyzupperpgas}-\ref{fig:rhormshave}).  The effect
of strong inhomogeneities on the emergent spectrum has been investigated in
the ordered shock train geometry produced in the initial nonlinear development
of the photon bubble instability \citep{dav04,beg06}.  At least in this
geometry, and at least
provided thermal absorption/emission dominates Compton scattering in the
energy exchange between radiation and matter, the high density
regions enhance the thermalization of radiation with matter.  This typically
produces a {\it softening} of the spectrum compared to what would emerge from an
atmosphere which is a horizontal average of the medium.
It remains to be seen which of these two opposing effects on the spectrum,
softening by inhomogeneities vs. hardening by magnetic support, will be
dominant, and how much effect Compton scattering can have in those regions
where the gas temperature rises well above the radiation temperature.

The very inhomogeneous and irregular nature of the photosphere is also
likely to reduce the degree of polarization of the emitted photons from
a scattering dominated atmosphere, as it breaks the plane parallel symmetry.
These inhomogeneities represent a physical realization of the ``rough
surface'' proposed and investigated by \cite{col90} in the context of
continuum polarization from active galactic nuclei.  As first suggested
by \citet{gne78}, Faraday depolarization may also be significant.  The
estimated rotation angle at the peak wavelength of a thermal spectrum
[i.e., the peak of $B_\lambda (T)$] is
$\simeq0.8\tau_TR^{1/2}$~radians.  Here $\tau_T$ is the Thomson optical
depth to the surface where the underlying polarization is imprinted;
we expect that because the opacity in these
disk atmospheres is primarily Thomson scattering, $\tau_T \simeq 1$.
The Faraday rotation is also proportional to $R$, the ratio of magnetic
pressure to radiation pressure, if we suppose that the component
of the field along the line of sight is a fixed and substantial
fraction of the total.
Using horizontal averages of the pressures, the highest total energy epoch in
the current simulation ($t=90$ orbits) has $R\sim7$ at the lower photosphere
and $R\sim3.5$
at the upper photosphere.  In both, the rotation angle is then
$\sim2$ radians per unit Thomson depth, large enough to
give rise to significant Faraday depolarization.

Magnetic support of the surface layers could also have implications
for X-ray reflection models that incorporate only gas and radiation pressure
in the hydrostatic equilibrium (e.g. \citealt{nay00}).  Most notably, the
reduced density that follows from magnetic support will make photoionization
relatively stronger.  It is also unclear that
the discontinuous, thermally stable structures that form in these models
would exist if magnetic fields dominate the hydrostatic support.
The right hand panel of Fig.~\ref{fig:rhofluct} shows that large density
inhomogeneities exist on the scale of a Thomson depth, and these may also
affect the reflection spectrum.  This effect
has been explored in a preliminary fashion by \citet{bal04} and \citet{bal05},
who performed radiative transfer calculations through one
dimensional inhomogeneous structures.  Density variations of only a factor
of a few were enough to produce significant changes in the reflection
spectrum \citep{bal04}, and we have more than that at the Rosseland mean
($\simeq$ Thomson) photosphere.  Of course, the inhomogeneities
themselves might be affected by heating by external radiation fields.
This topic is clearly one that deserves further investigation.

\section{Summary}

    Hydrostatic equilibrium is often thought of in terms of a balance between
gravity and any of three sorts of pressures: gas, radiation, and magnetic.
However, we find that in the upper, magnetically-dominated layers of this
disk segment, magnetic tension forces are almost as strong as magnetic
pressure forces, at least in a volume-averaged sense.  This is clearly
because the nonlinear development of the transverse Parker instability is
controlling the dynamics of the magnetic field in these upper layers.  In
a mass-averaged sense, however, magnetic pressure is still the dominant
source of vertical support.

In contrast to the Parker instability, there is very little evidence for
photon bubbles, even though their predicted linear growth rates are almost as
large as the Parker growth rates, and the relevant unstable wavelengths are
resolved.

Strong density inhomogeneities are present in the region between the effective
(thermalization) and Rosseland mean ($\simeq$ scattering) photospheres.  This
was also true in the gas pressure dominated simulation of \citet{hir06}, and
these inhomogeneities are almost certainly the result of the large magnetic
forces in the surface layers.
Provided Compton scattering is less important than true absorption
processes in thermally coupling photons with matter, these inhomogeneities
are likely to enhance thermalization and therefore soften the emergent spectrum.
This spectral softening mechanism may counteract to some extent the
reduction in thermalization, and
therefore the spectral hardening, caused by the lower average densities
that are produced by magnetic support of the atmosphere.
Inhomogeneities are likely to reduce the polarization of the emergent photons,
and the magnetic fields in the atmosphere will reduce it still further
through Faraday depolarization.  Finally, the inhomogeneities we see at the
scattering photosphere appear to be strong enough to affect X-ray reflection
spectra.

More work needs to be done to elaborate on these conclusions.
Given that the size of the linearly unstable
Parker wavelengths is comparable to the azimuthal size of the box, it would
be worthwhile in future to run simulations with larger azimuthal domains to
see if this alters the magnetic equilibrium in the upper layers.
Unfortunately, these simulations are expensive, so performing a full
parameter survey is difficult.

It is not completely clear why we have failed to find evidence for photon
bubbles in the present simulation.  The accessible growing wavelengths are
not too far below the turnover wavelength, and increasing the dynamic range
of the simulation by decreasing the cell size would enable us to access even
faster growing wavelengths which might then be better able to compete
with the Parker instability.  However, to fully understand which instability
should dominate requires, at the very least, a linear
instability analysis of an equilibrium with both magnetic pressure support
and a diffusive radiation flux.  Such an equilibrium would then be more
representative of the conditions we are finding in the simulation, and
would be vulnerable to both the Parker and photon bubble instabilities.
It could be that these
two instabilities couple together in nontrivial ways, although their physical
driving mechanisms are quite different.  The transverse Parker instability
turns a slow magnetosonic wave into an exponentially growing disturbance if
magnetic pressure gradients are large enough.  The fastest growing photon bubble
instability is an overstability of, again, a slow magnetosonic wave, but is
driven by the background radiation pressure.

Whether the emergent spectrum is harder due to magnetic support or softer
due to density inhomogeneities when compared to plane parallel atmosphere
models that neglect magnetic support will require a three dimensional radiative
transfer calculation through the structures seen in the simulation.  Monte
Carlo calculations may be the way to approach this problem
(e.g. \citealt{dav04}), and these could also include polarization with
Faraday rotation \citep{ago96}.  It might also be possible to calculate
X-ray reflection in a similar fashion, though one would have to include
the resulting photoionization self-consistently.

It remains to be seen what will happen when
radiation pressure becomes dominant at the midplane.  Photon bubbles should
have even stronger growth rates in that regime, and may be more relevant.
However, they will also be more difficult to resolve in a simulation because
their characteristic wavelengths are only of order the gas pressure scale
height.  We expect to be able to report on simulations with large radiation
pressure in the near future.

We would like to thank Mitch Begelman, Shane Davis, Phil Marshall, Greg
Shields, Neal Turner, Tommaso Treu, and Ellen Zweibel for very useful
discussions.  We are also especially grateful
to Jim Stone for comments and insights that greatly improved this paper,
as well as for developing the simulation code we used.
This work was supported in part by NSF Grants AST-0307657
(OB) and AST-0507455 (JHK).  The numerical simulations were carried out on
the SX8 at the Yukawa Institute for Theoretical Physics of Kyoto University
and the VPP5000 at the Center for Computational Astrophysics of the National
Astronomical Observatory of Japan.

\appendix

\section{Appendix: Simplified Derivation of Parker Instability in a Radiating
Medium}

Assuming the gas and radiation are locked to the same temperature and that
the medium is optically thick, the equations of radiation magnetohydrodynamics
are
\begin{equation}
{\partial\rho\over\partial t}+\nabla\cdot(\rho{\bf v})=0,
\end{equation}
\begin{equation}
\rho\left({\partial{\bf v}\over\partial t}+{\bf v}\cdot\nabla{\bf v}\right)
=-\nabla p_{\rm g}+\rho{\bf g}+{1\over4\pi}{\bf B}\cdot\nabla{\bf B}-
{1\over8\pi}\nabla B^2+{\kappa_{\rm F}\rho\over c}{\bf F},
\end{equation}
\begin{equation}
{\partial\over\partial t}(e+E)+{\bf v}\cdot\nabla(e+E)+\left(\gamma e+{4\over3}
E\right)\nabla\cdot{\bf v}=-\nabla\cdot{\bf F},
\label{eq:energy}
\end{equation}
\begin{equation}
{\bf F}=-{c\over3\kappa_{\rm F}\rho}\nabla E,
\label{eq:raddiff}
\end{equation}
and
\begin{equation}
{\partial{\bf B}\over\partial t}=\nabla\times({\bf v}\times{\bf B}),
\end{equation}
where
\begin{equation}
e={p_{\rm g}\over(\gamma-1)},\,\,\,\,\,
p_{\rm g}={\rho k_{\rm B}T\over\mu},\,\,\,\,\,{\rm and}
\,\,\,\,\,E=aT^4.
\end{equation}

We adopt a simplified, static equilibrium structure in which the gravitational
acceleration ${\bf g}=-g\hat{\bf z}$ is constant.  Including the linear
increase of $g$ with height merely complicates the form of the unstable
eigenfunctions and is not really justified here given the additional
simplifications we are making.  These include assuming that the background
medium is isothermal and that the background magnetic energy density is
proportional to the density.  These assumptions imply that the isothermal
sound speed in the gas $c_{\rm i}$ and the Alfv\'en speed $v_{\rm A}$ are
constants, which greatly simplifies the analysis.  A more serious consequence
of these assumptions is that the background radiation flux ${\bf F}$ vanishes,
and does not help support the medium against gravity.  Because of this, we
immediately lose the driving that causes photon bubbles.  Nevertheless, we
still hope to capture the basic growth rates and wavelength scales of the
Parker instability in the magnetically dominated upper layers of our
simulation.  Finally, we assume that the background magnetic field is
horizontal: ${\bf B}=B(z)\hat{\bf y}$.
 
With all these assumptions, hydrostatic
equilibrium requires that the background density satisfies
\begin{equation}
\rho=\rho_0\exp\left(-{z\over H}\right),
\end{equation}
where
\begin{equation}
H={2c_{\rm i}^2+v_{\rm A}^2\over2g}.
\end{equation}

Linearizing equations (\ref{eq:energy}) and (\ref{eq:raddiff}) about our
background, and assuming that our disturbances are at short enough wavelengths
that the rapid diffusion limit of equation (\ref{eq:lamr}) is satisfied
immediately gives us $\delta T=0$.  Assuming the perturbations have a
spacetime dependence of $f(z)\exp[i(k_xx+k_yy-\omega t)]$, and eliminating
the magnetic field and pressure perturbations using the flux-freezing equation
and equation of state, the linearized continuity and momentum equations become
\begin{equation}
-i\omega{\delta\rho\over\rho}+ik_x\delta v_x+ik_y\delta v_y-
{\delta v_z\over H}+{\partial \delta v_z\over\partial z}=0,
\label{eq:dcont}
\end{equation}
\begin{equation}
(\omega^2-k_x^2v_{\rm A}^2-k_y^2v_{\rm A}^2)\delta v_x-{ik_xv_{\rm A}^2\over2H}
\delta v_z+ik_xv_{\rm A}^2{\partial\delta v_z\over\partial z}-\omega k_x
c_{\rm i}^2{\delta\rho\over\rho}=0,
\label{eq:dvx}
\end{equation}
\begin{equation}
\omega^2\delta v_y-{ik_yv_{\rm A}^2\over2H}\delta v_z-\omega k_yc_{\rm i}^2
{\delta\rho\over\rho}=0,
\label{eq:dvy}
\end{equation}
and
\begin{equation}
\left(\omega^2-k_y^2v_{\rm A}^2+{v_{\rm A}^2\over2H^2}\right)\delta v_z
-{3v_{\rm A}^2\over2H}{\partial\delta v_z\over\partial z}+v_{\rm A}^2
{\partial^2\delta v_z\over\partial z^2}+{i\omega v_{\rm A}^2\over2H}
{\delta\rho\over\rho}+i\omega c_{\rm i}^2{\partial\over\partial z}
\left({\delta\rho\over\rho}\right)-{ik_xv_{\rm A}^2\over H}\delta v_x
+ik_xv_{\rm A}^2{\partial\delta v_x\over\partial z}=0.
\label{eq:dvz}
\end{equation}

We now adopt the {\it ansatz}
that the $z$-dependence of the velocity perturbations and relative density
perturbation is $\exp[ik_zz+z/(2H)]$, where $k_z$ is a real number representing
the wave vector in the vertical direction.  The $\exp[z/(2H)]$ factor implies
that $\rho|\delta v^2|$ is constant with height, thereby guaranteeing
energy conservation for propagating waves (see e.g. section 53 of
\citealt{mih84}).  We then obtain the dispersion relation
\begin{eqnarray}
&&\omega^6-\left[\left(k^2+{1\over4H^2}\right)(c_{\rm i}^2+v_{\rm A}^2)+k_y^2
v_{\rm A}^2\right]\omega^4+v_{\rm A}^2\left(2c_{\rm i}^2+v_{\rm A}^2\right)
\left(k_y^2k^2+{k_x^2\over4H^2}\right)\omega^2\cr
&&+k_y^2v_{\rm A}^4\left[-k^2k_y^2c_{\rm i}^2+{k_x^2\over4H^2}
\left(2c_{\rm i}^2+v_{\rm A}^2\right)+
{k_y^2\over4H^2}(c_{\rm i}^2+v_{\rm A}^2)\right]=0,
\label{eq:parkerdisp}
\end{eqnarray}
where $k^2\equiv (k_x^2+k_y^2+k_z^2)^{1/2}$ is the total wavenumber of the
perturbation.  Equation (\ref{eq:parkerdisp}) is identical to the original
dispersion relation (I.19) of \citet{par67}, restricted to isothermal
perturbations.

Instability only exists if $k_y\ne0$, and when it does, maximum growth occurs
for horizontal wavenumbers ($k_z=0$).   In this case, instability occurs only
for wavelengths $\lambda_y\equiv2\pi/k_y$ satisfying
\begin{equation}
\lambda_y>\lambda_{\rm Parker}\equiv\cases{
{2\pi c_{\rm i}(2c_i^2+v_{\rm A}^2)\over g(c_{\rm i}^2+v_{\rm A}^2)^{1/2}}
& if $k_x=0$,\cr
{2\pi c_{\rm i}(2c_{\rm i}^2+v_{\rm A}^2)^{1/2}\over g}
& if $k_x\rightarrow\infty$.
}
\label{eq:lamparker}
\end{equation}
Note that if the magnetic energy density is dominant over gas pressure,
$\lambda_{\rm Parker}\simeq2\pi c_{\rm i}v_{\rm A}/g$.  The wavelength
$\lambda_y$ for maximum growth occurs at
\begin{equation}
\lambda_{\rm max}=\cases{
{2\pi(2c_{\rm i}^2+v_{\rm A}^2)|v_{\rm A}^2-c_{\rm i}^2|c_{\rm i}^{1/2}
\over g\{(c_{\rm i}^2+v_{\rm A}^2)[2^{1/2}(c_{\rm i}^2
+v_{\rm A}^2)^{3/2}-c_{\rm i}(c_{\rm i}^2+3v_{\rm A}^2)]\}^{1/2}}
& if $k_x=0$,\cr
{2\pi v_{\rm A}^2(2c_{\rm i}^2+v_{\rm A}^2)^{1/2}c_{\rm i}^{1/2}\over g
[2^{1/2}(2c_{\rm i}^2+v_{\rm A}^2)^{3/2}
-(4c_{\rm i}^2+3v_{\rm A}^2)c_{\rm i}]^{1/2}}
& if $k_x\rightarrow\infty$.
}
\label{eq:lammax}
\end{equation}
In the limit of strong magnetic fields, $\lambda_{\rm max}\simeq2\pi v_{\rm A}
(v_{\rm A}c_{\rm i})^{1/2}/(2^{1/4} g)$.  The corresponding maximum growth
rate is
\begin{equation}
\gamma_{\rm max}=\cases{
{gv_{\rm A}(c_{\rm i}^2+v_{\rm A}^2)^{1/2}[3c_{\rm i}^2+
v_{\rm A}^2-2^{3/2}(c_{\rm i}^2+v_{\rm A}^2)^{1/2}c_{\rm i}]^{1/2}\over
(2c_{\rm i}^2+v_{\rm A}^2)|v_{\rm A}^2-c_{\rm i}^2|}.
& if $k_x=0$,\cr
{g[4c_{\rm i}^2+v_{\rm A}^2-2^{3/2}(2c_{\rm i}^2+v_{\rm A}^2)^{1/2}c_{\rm i}
]^{1/2}\over v_{\rm A}(2c_{\rm i}^2+v_{\rm A}^2)^{1/2}}
& if $k_x\rightarrow\infty$.
}
\label{eq:gammamax}
\end{equation}
For strong magnetic fields, this becomes simply
$\gamma_{\rm max}\simeq g/v_{\rm A}$.


\begin{figure}
\epsscale{1.0}
\plotone{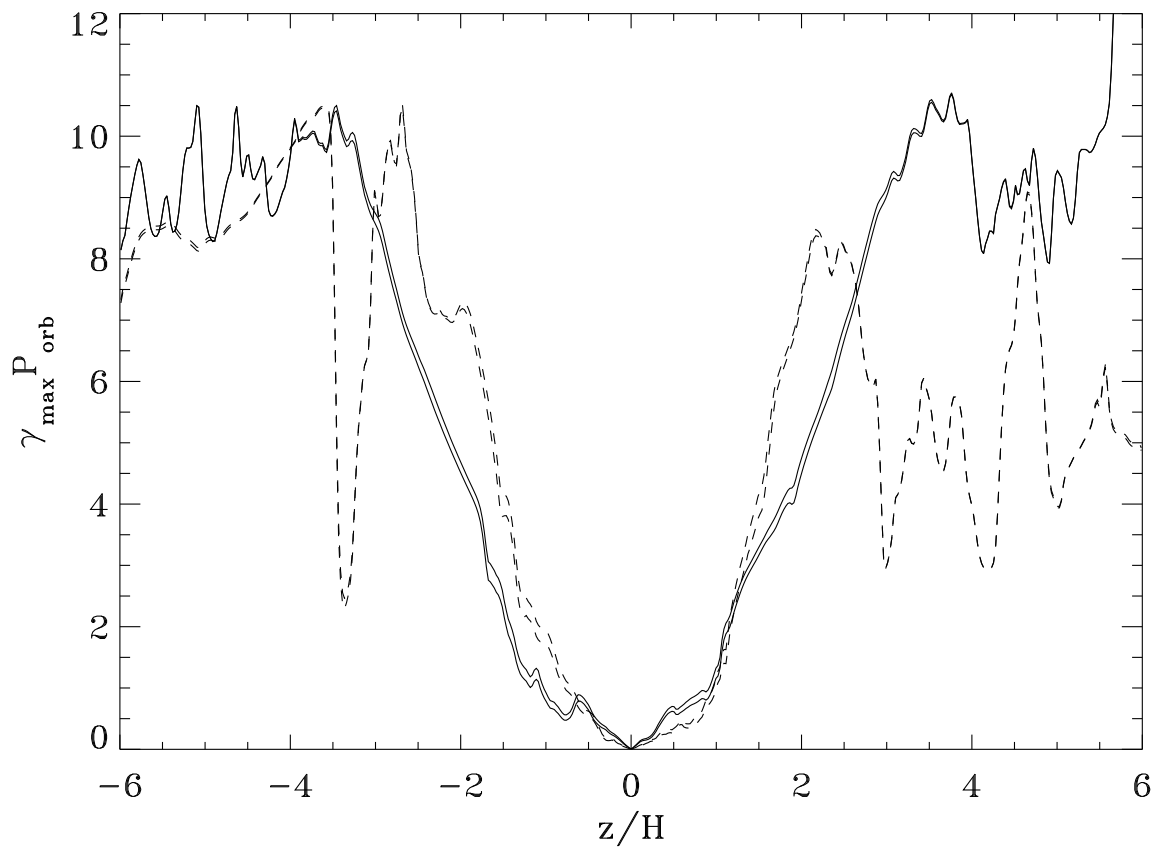}
\caption{Maximum Parker instability growth rate $\gamma_{\rm max}$ from
equation (\ref{eq:gammamax}),
scaled with the
orbital period $P_{\rm orb}=2\pi/\Omega$, as a function of height in the
box at $t=90$ orbits (solid) and $t=150$ orbits (dashed).  The two nearly
identical curves for each epoch correspond to the $k_x=0$ and $\infty$
limits.}
\label{fig:parkergrowthrate}
\end{figure}

\begin{figure}
\epsscale{1.0}
\plottwo{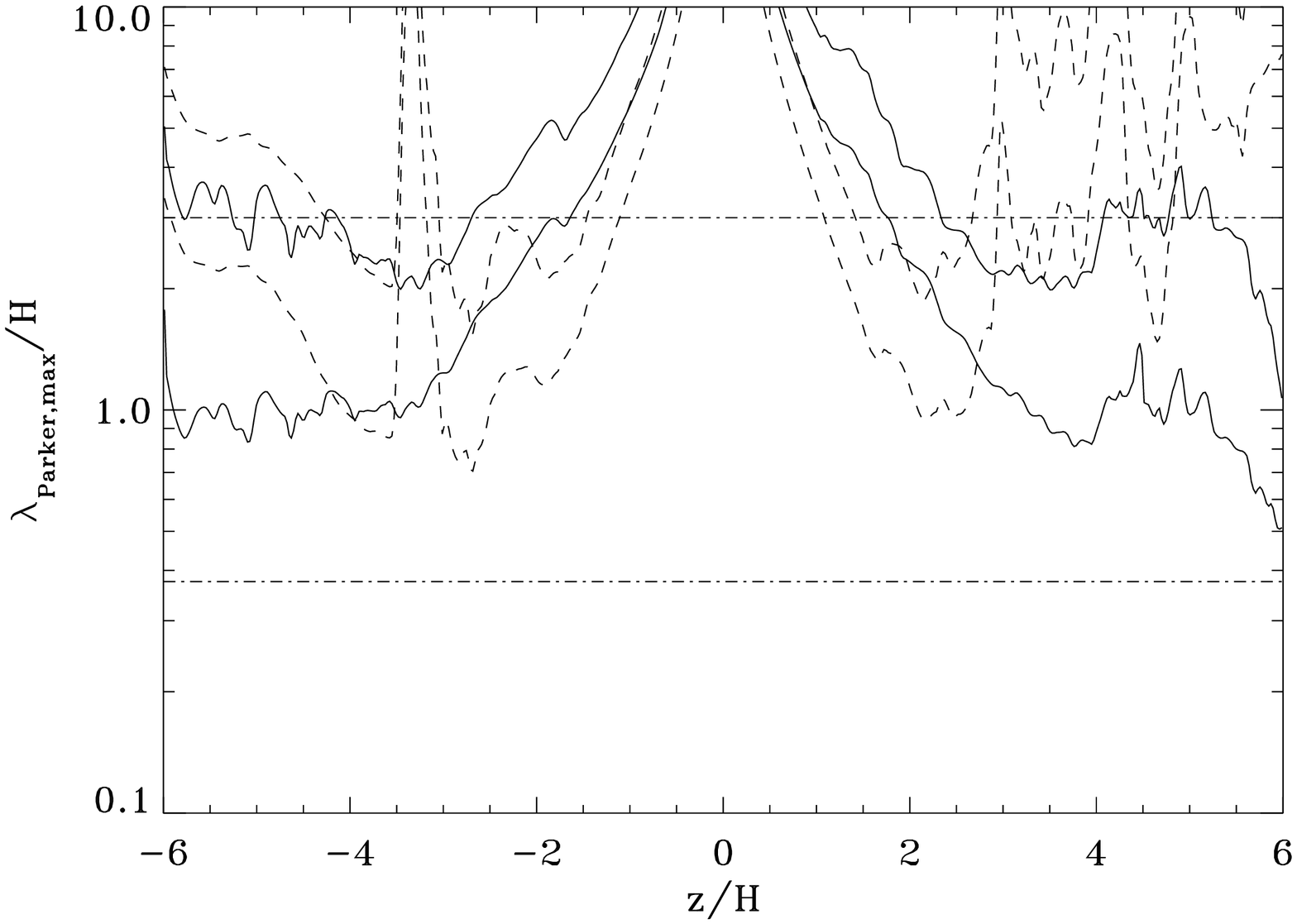}{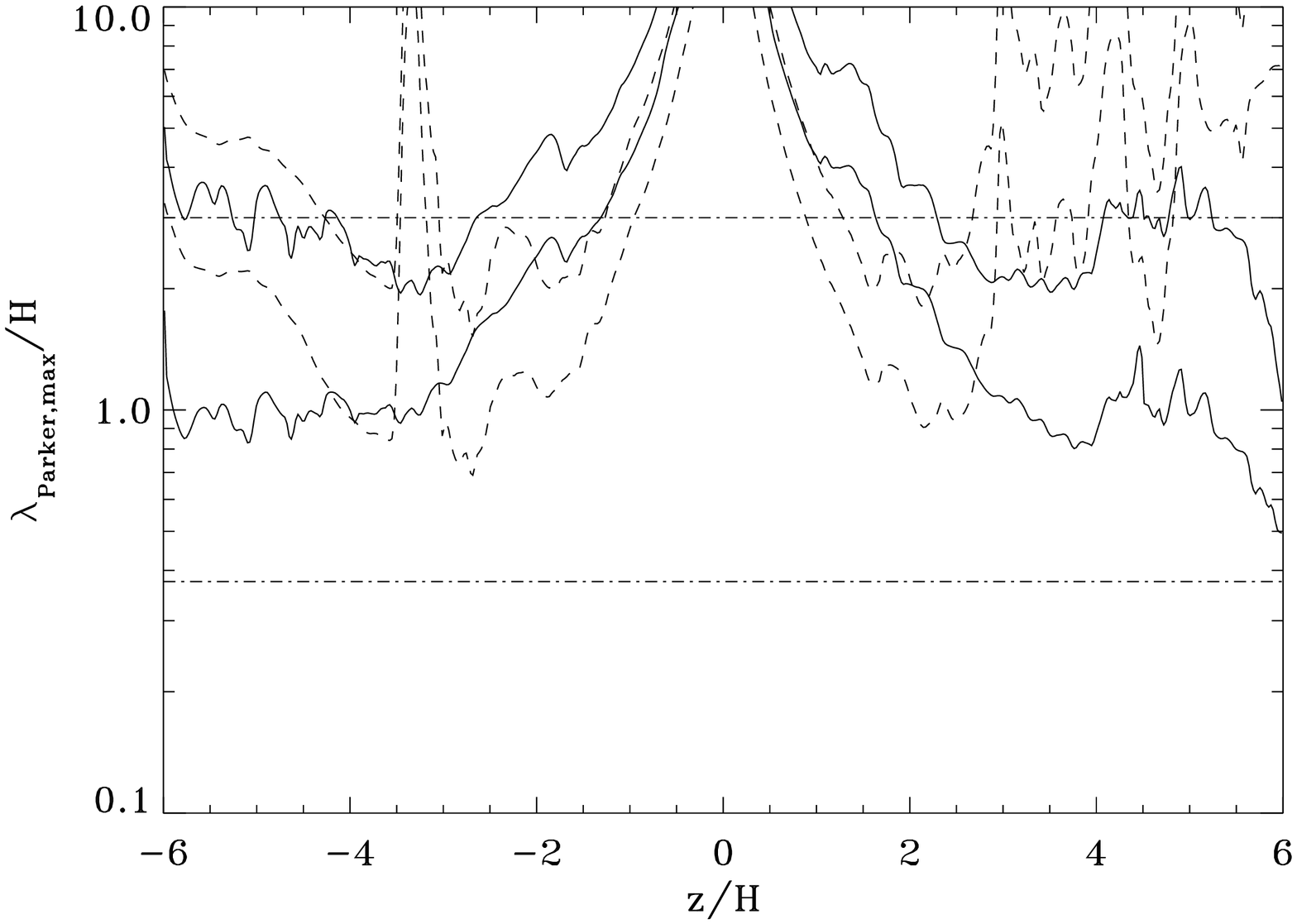}
\caption{Minimum Parker instability wavelength from equation
(\ref{eq:lamparker}) (lower of each pair of curves)
and wavelength for maximal Parker growth rate from equation (\ref{eq:lammax})
(upper of each pair of curves)
as a function of height in the box at $t=90$ orbits (solid curve pair) and
$t=150$ orbits (dashed curve pair).
Eight cell sizes and the length of the box in the azimuthal direction are
indicated again by the lower and upper horizontal dashed lines, respectively.
The left hand panel corresponds to $k_x=0$, while the right hand panel
corresponds to $k_x\rightarrow\infty$.  The dependence on $k_x$ is very
weak.}
\label{fig:rlammax}
\end{figure}

\begin{figure} 
\epsscale{1.0}
\plotone{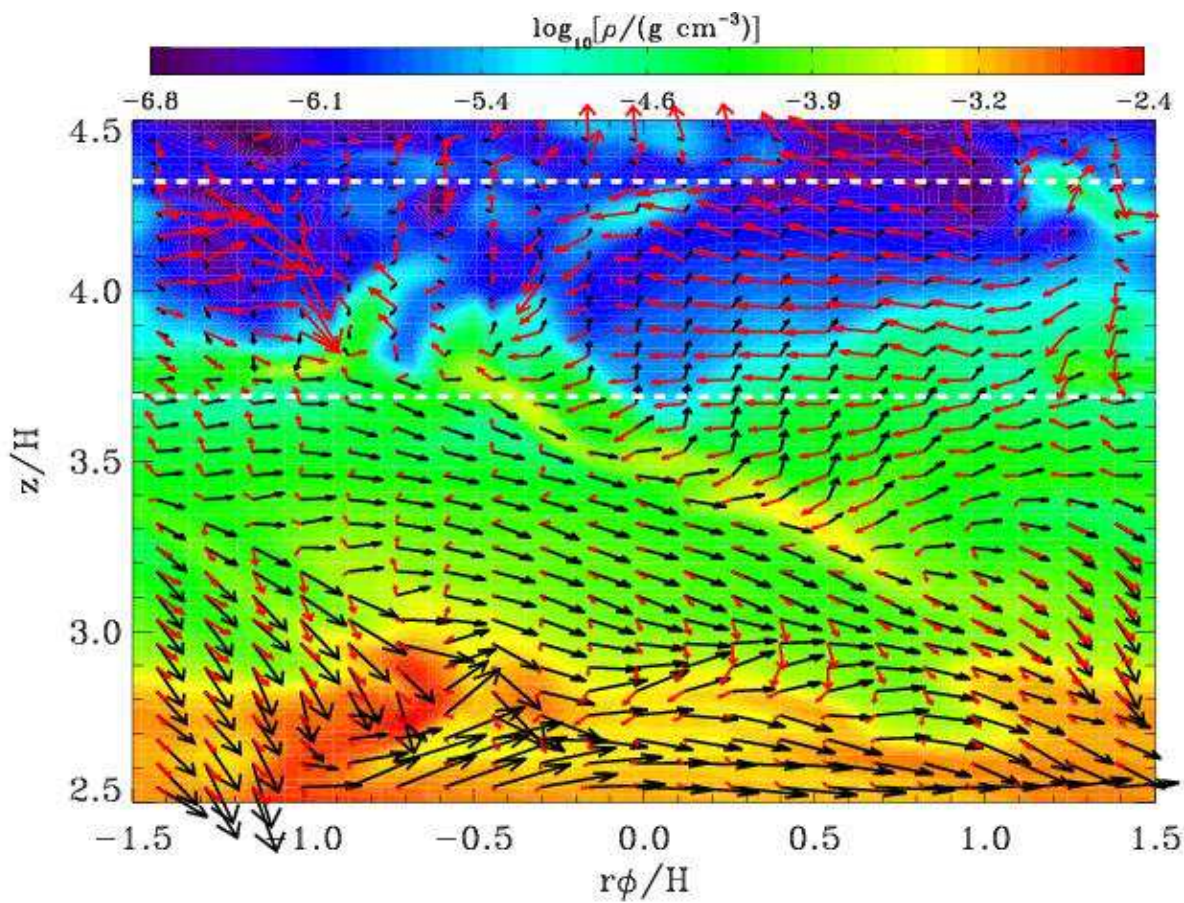}
\caption{Density in regions under the horizontally-averaged upper photosphere
(dashed line) in a fixed radial slice near the middle of the box at
$t=90$~orbits.  The arrows show the projections of magnetic field vectors
(black) and velocity vectors relative to the background shear flow (red)
in this radial slice, computed at the position of the tail of each arrow.
Horizontal dashed lines indicate the positions of the Rosseland mean
photosphere (upper) and effective photosphere (lower), computed from the
horizontally averaged structure.
\label{fig:rhoyzupper090}}
\end{figure}

\begin{figure}
\epsscale{1.0}
\plotone{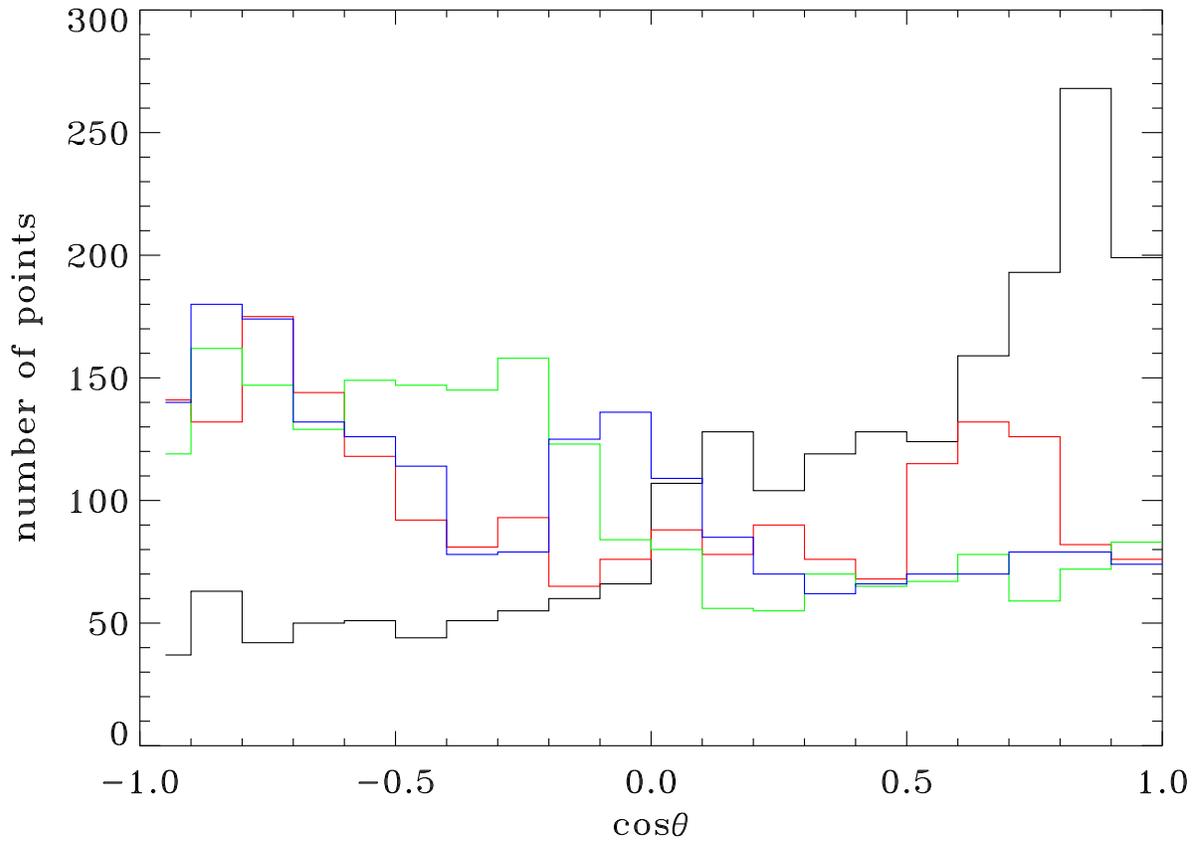}
\caption{Distribution of cosines of the angle $\theta$ between the
perturbed radiation
flux vector and the fluid velocity relative to the background shear flow, over
horizontal slices in the simulation at $t=90$ orbits.  Different histograms
correspond to different heights:  $+2.4H$ (black), $+2.9H$ (red), $+3.3H$
(green), and $+3.8H$ (blue).
\label{fig:pbidiagnostic090}}
\end{figure}

\begin{figure}
\epsscale{1.0}
\plottwo{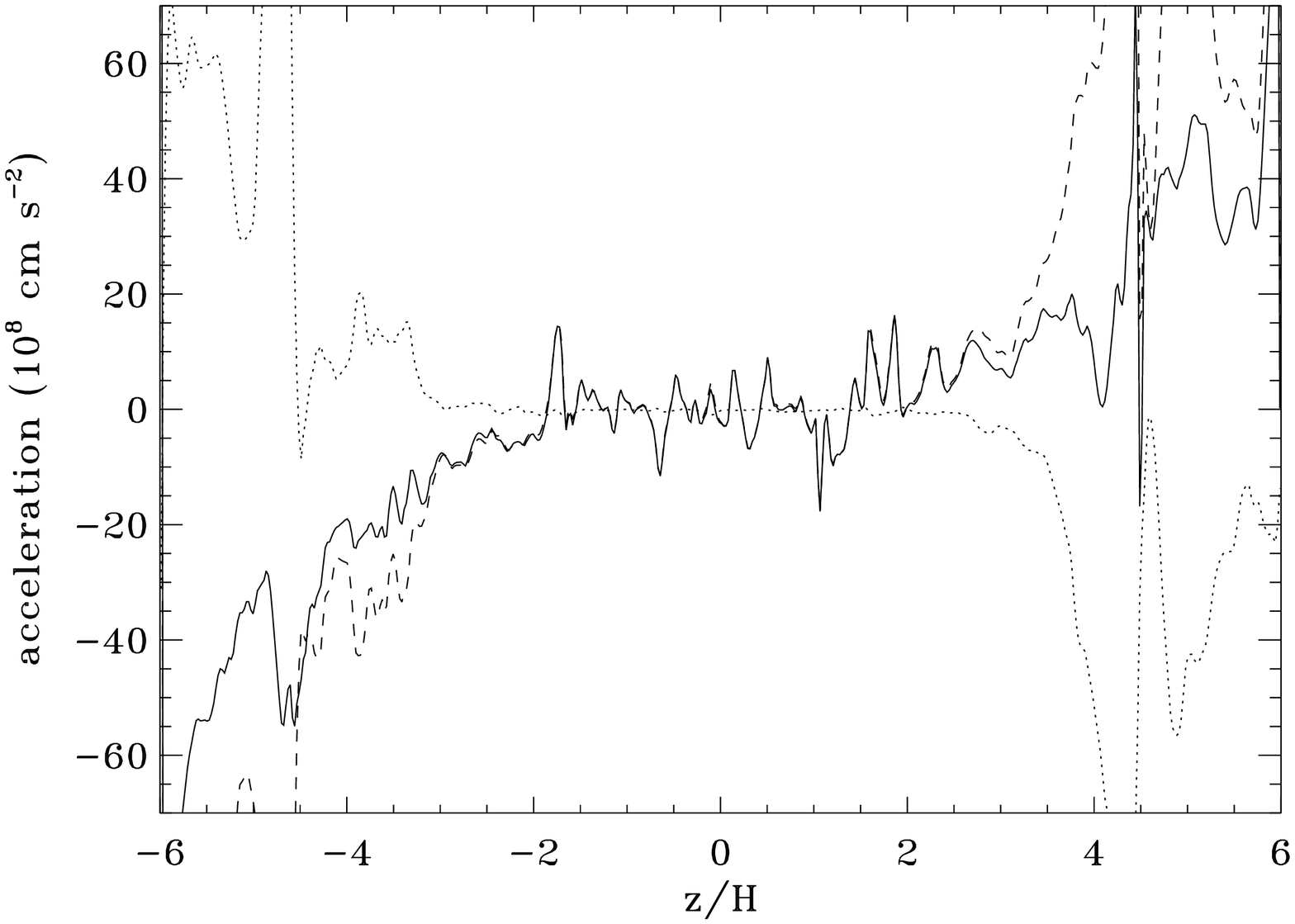}{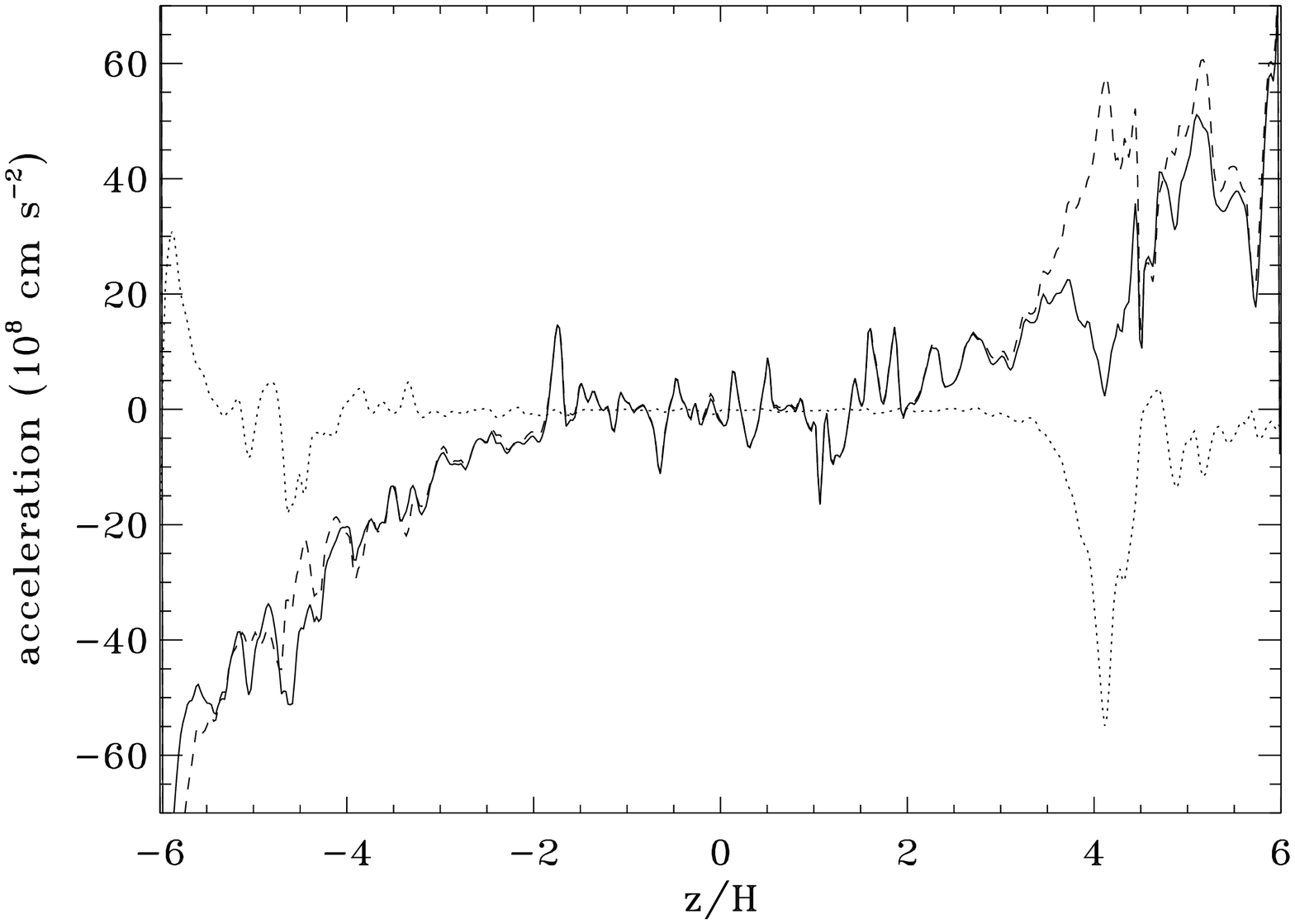}
\caption{Spatially horizontally averaged (left) and mass-weighted horizontally
averaged (right) contributions to the outward vertical
accelerations from magnetic pressure gradients (dashed) and magnetic
tension (dotted), at $t=90$ orbits.  The solid curves show the sum of these
magnetic accelerations.}
\label{fig:vertaccmag}
\end{figure}

\begin{figure}
\epsscale{1.0}
\plottwo{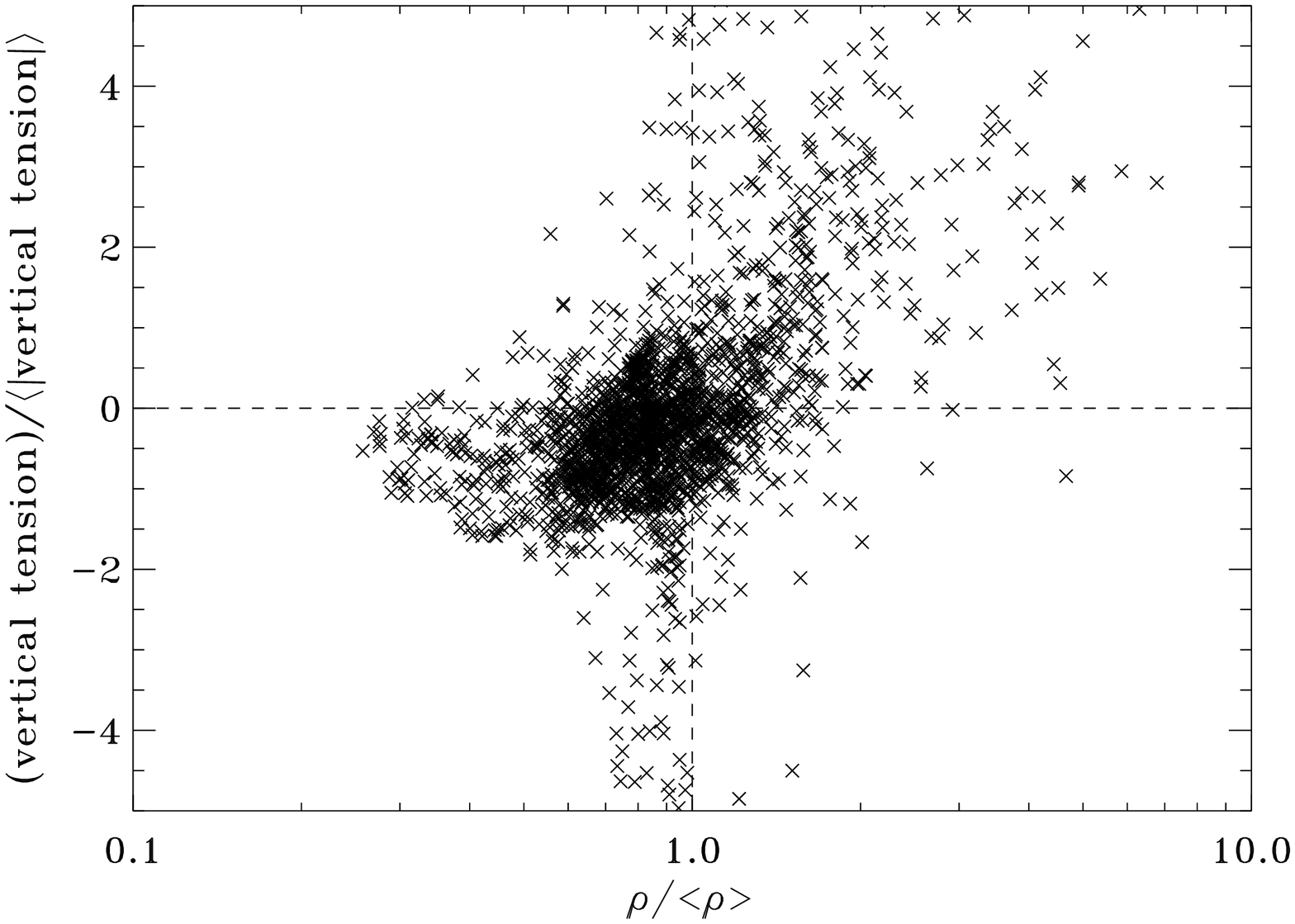}{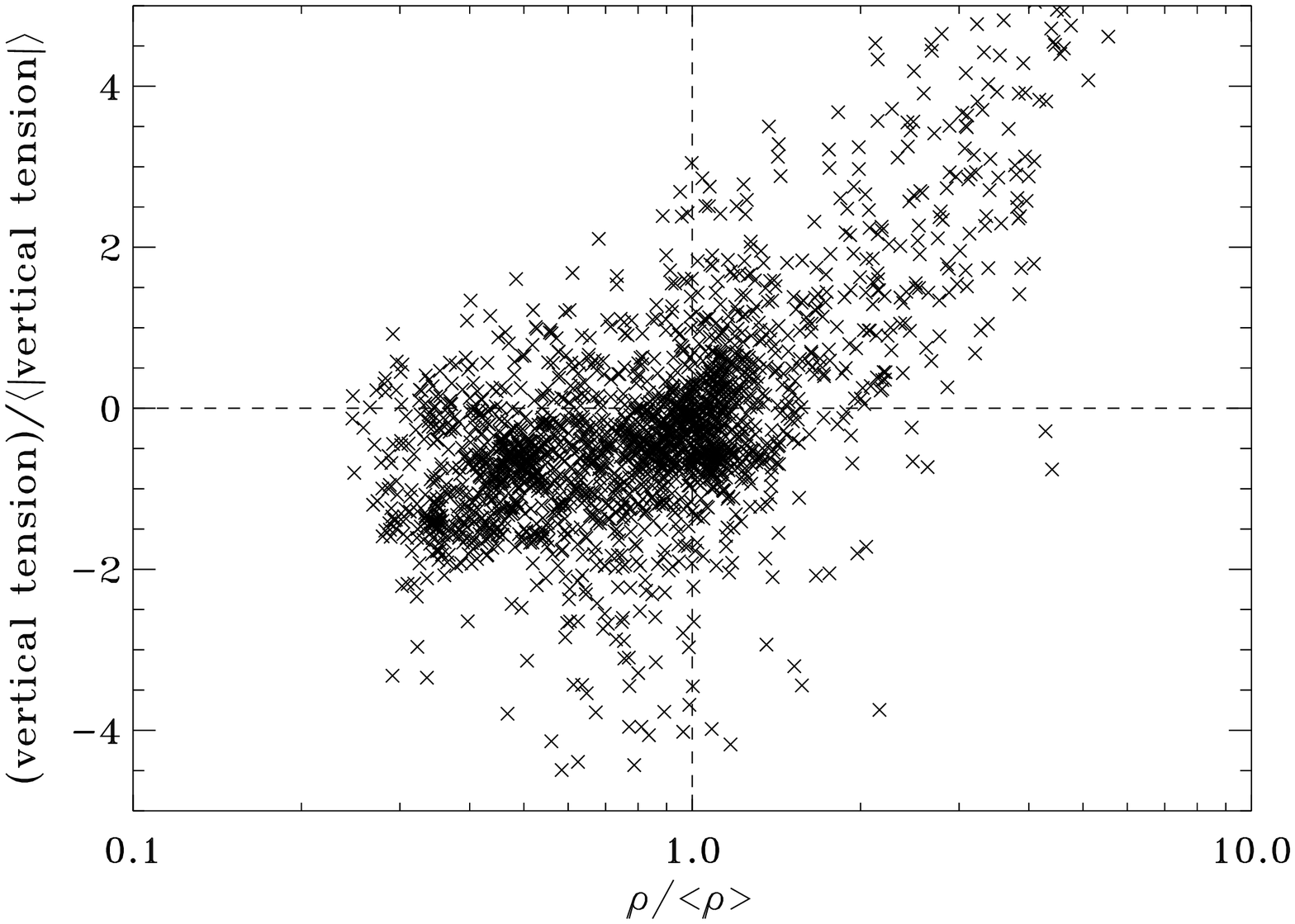}
\caption{Vertical magnetic tension force per unit volume vs. density at each
grid zone in a $z=2.75H$ horizontal slice through the simulation at $t=90$
orbits (left) and in a $z=2H$ horizontal slice through the simulation at
$t=150$ orbits (right).  The tension has been scaled by the average magnitude
of the tension
force in this horizontal slice, and the density has been scaled by the
average density.
\label{fig:tensionvsrho}}
\end{figure}

\begin{figure}
\epsscale{1.0}
\plotone{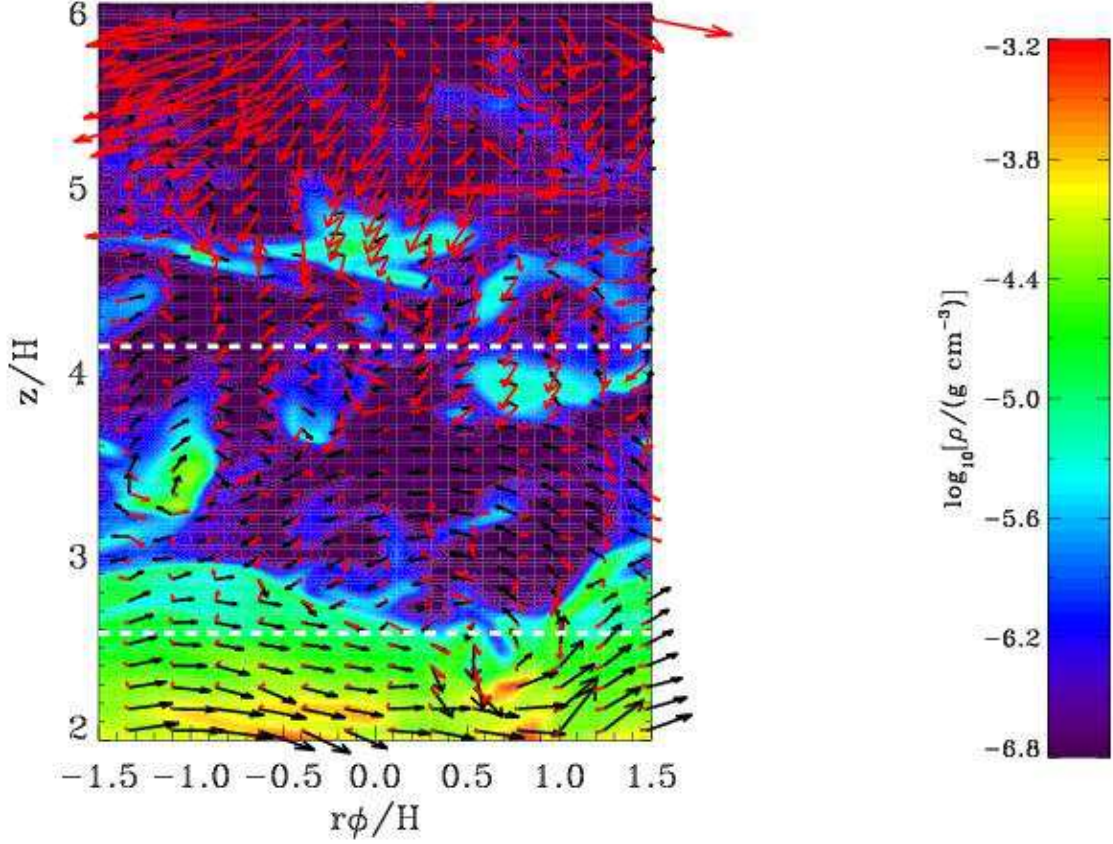}
\caption{Density in the upper layers ($2H<z<6H$) in a fixed radial
slice near the middle of the box at time 150 orbits.  The arrows show the
projections of magnetic field vectors (black) and velocity vectors
relative to the background shear flow (red) into this radial slice, computed
at the position of the tail of each arrow.  The Rosseland mean and
effective photospheres of the horizontally averaged structure at this time are
indicated by the upper and lower dashed lines, respectively.
\label{fig:rhoyzupper150}}
\end{figure}

\begin{figure}
\epsscale{1.0}
\plottwo{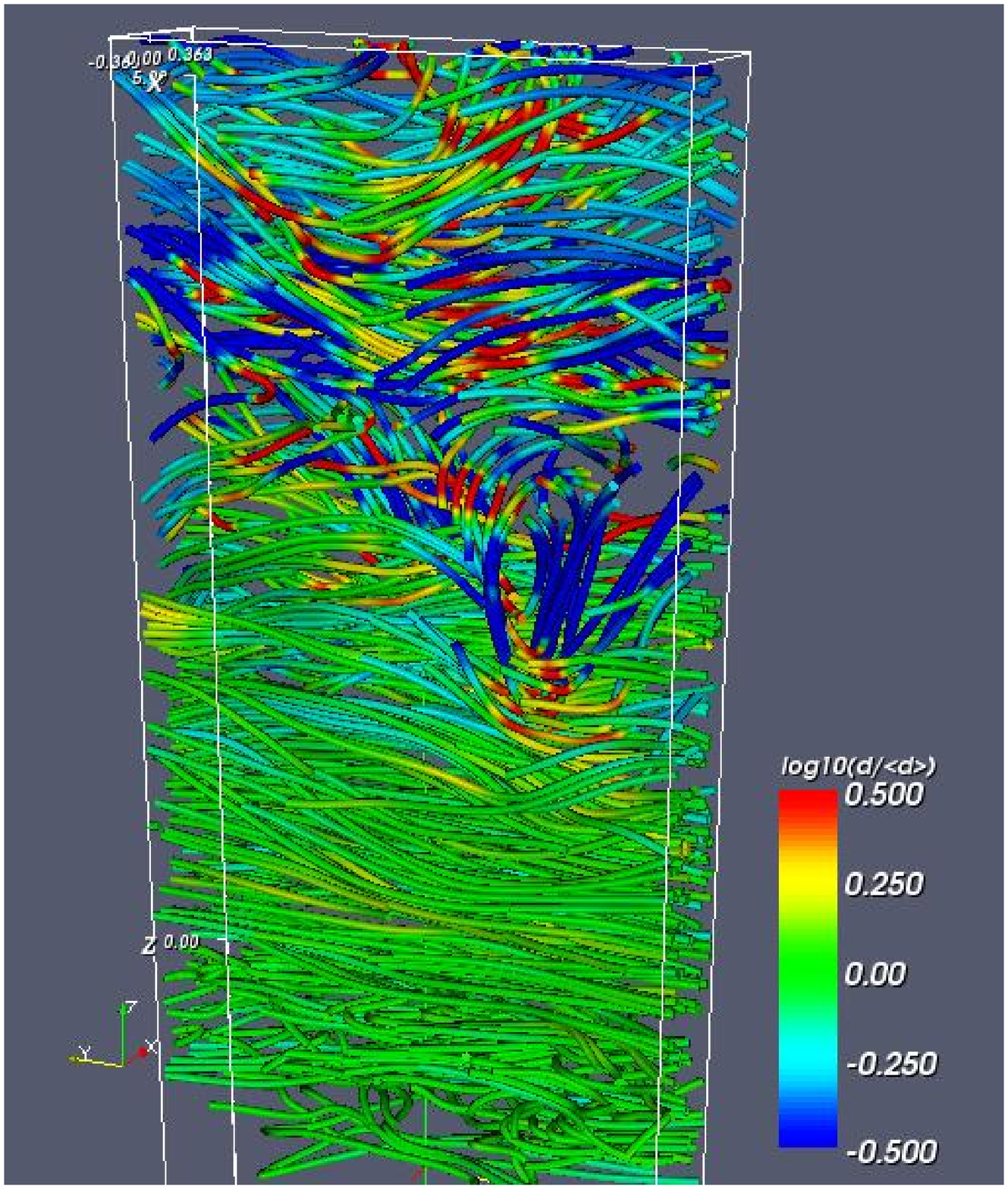}{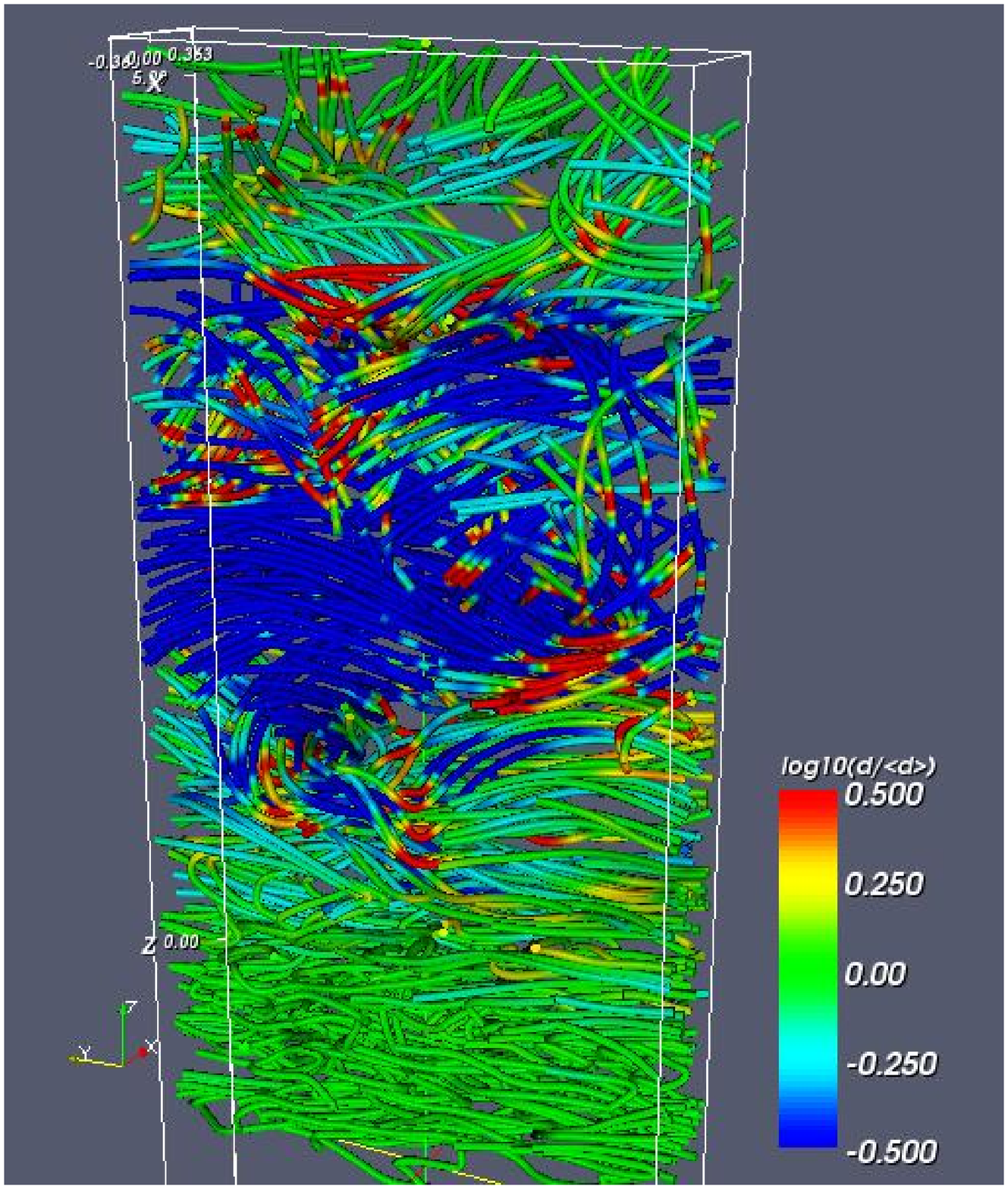}
\caption{Magnetic field lines in the simulation domain at 90 orbits (left)
and 150 orbits (right).  The lines are color coded with the local fluid
density scaled by the horizontally averaged density at the same height.}
\label{fig:fieldlines}
\end{figure}

\begin{figure}
\epsscale{1.0}
\plottwo{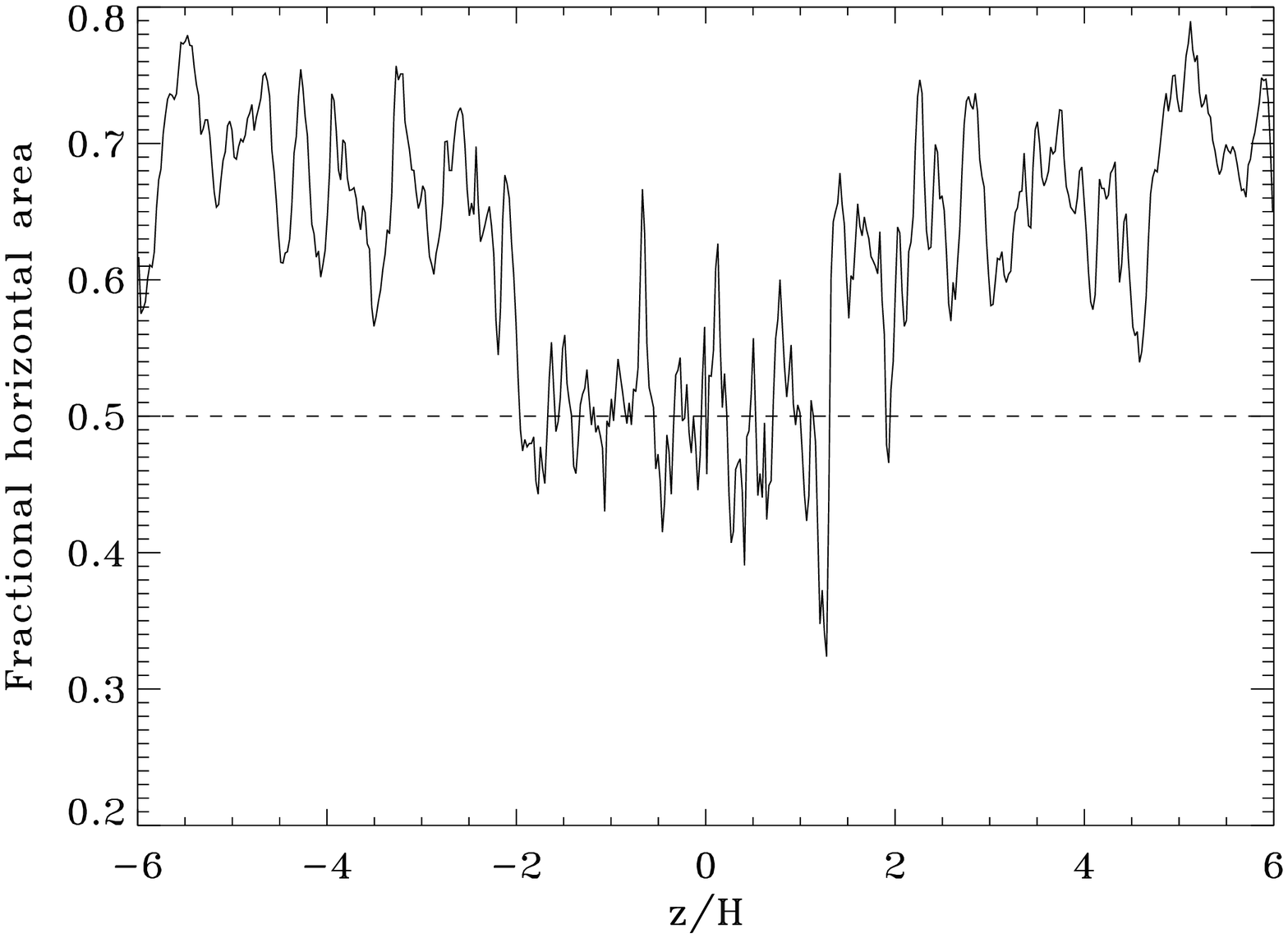}{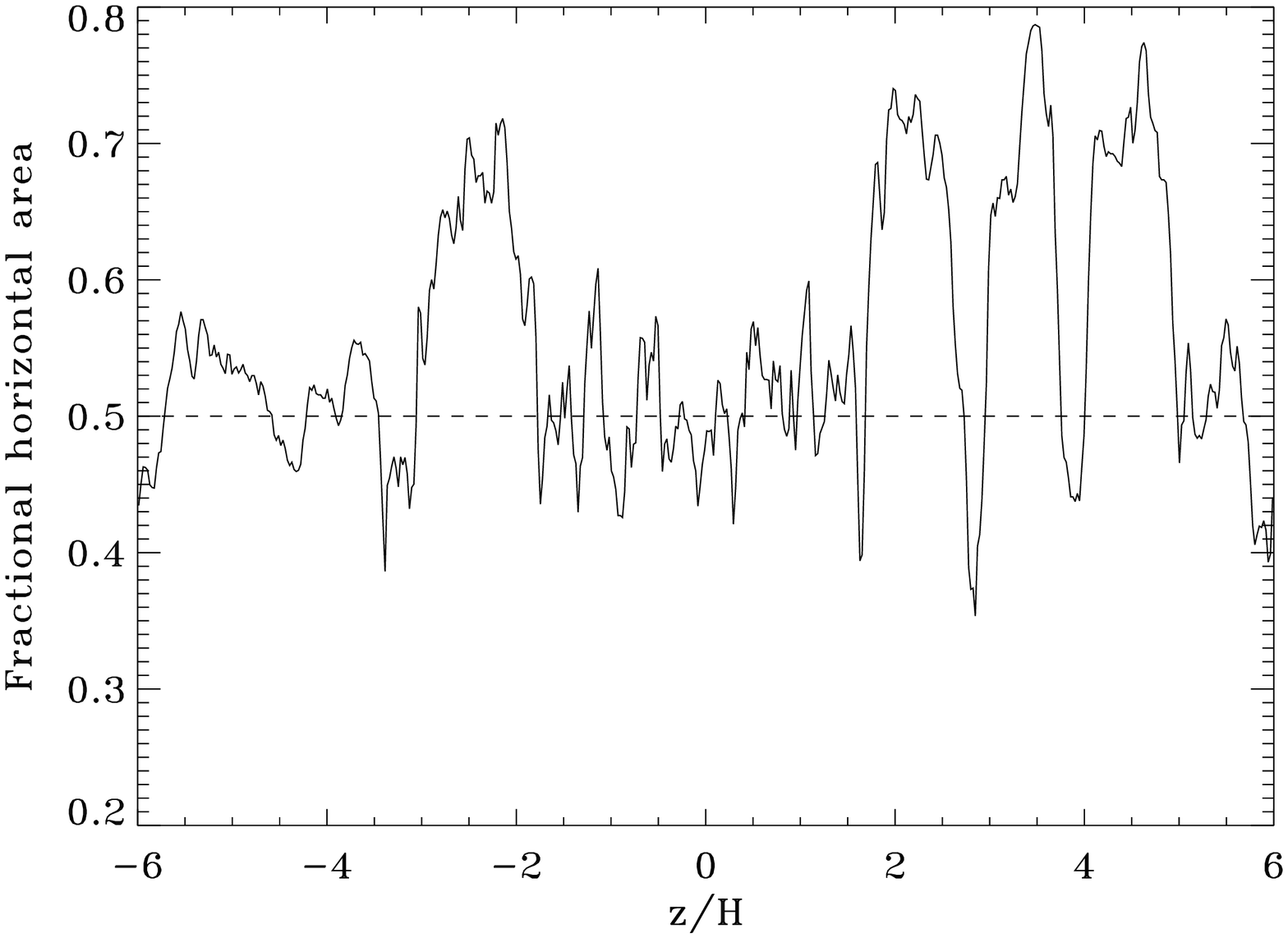}
\caption{The fractional horizontal area over which higher (lower) than
average densities are associated with vertically inward (outward) magnetic
tension forces, as a function of height in the simulation.  The left panel is
for $t=90$ orbits and the right panel is for $t=150$ orbits.
\label{fig:tensionrhostats}}
\end{figure}

\begin{figure}
\epsscale{1.0}
\plotone{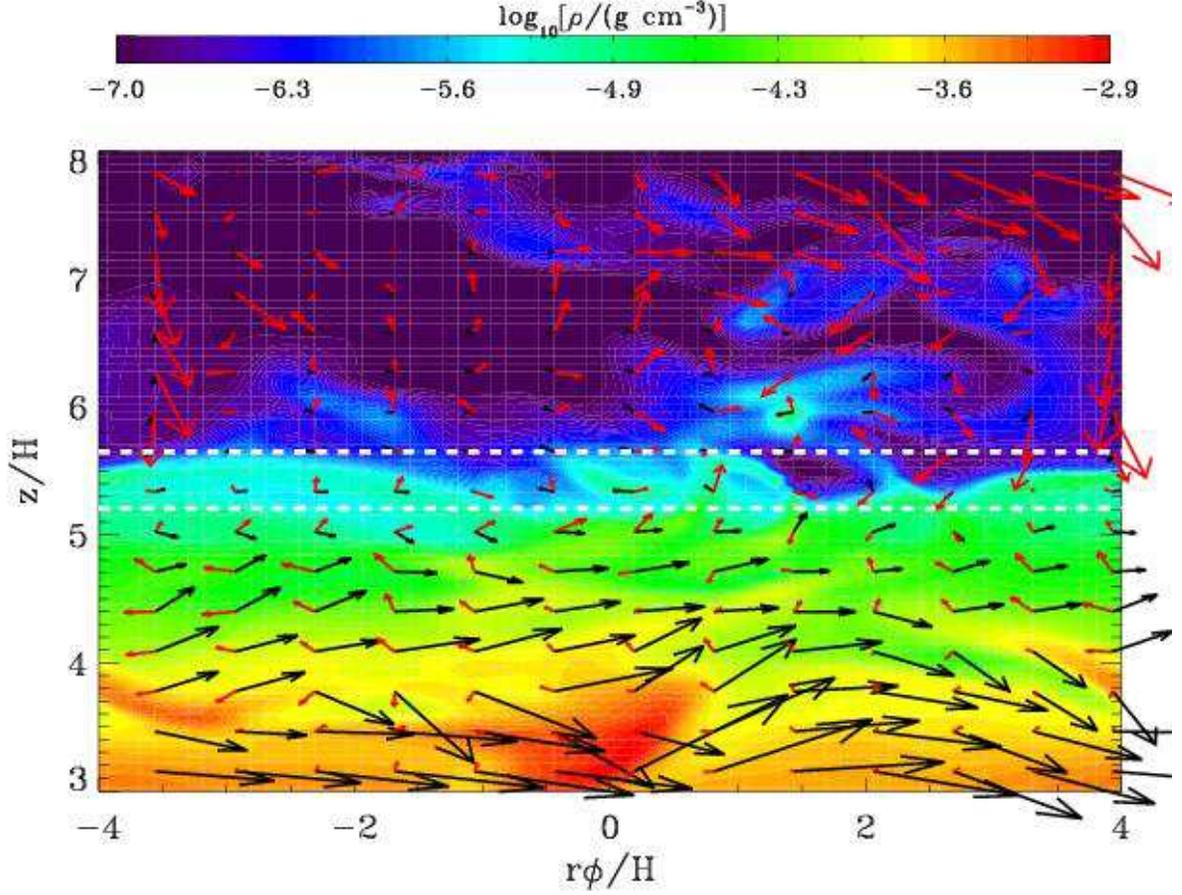}
\caption{Density in the uppermost surface layers ($3H<z<8H$) in a fixed radial
slice near the middle of the box at time 60 orbits from the gas pressure
dominated simulation of \citet{hir06}.  The arrows show the
projections of magnetic field vectors (black) and velocity vectors relative
to the background shear flow (red)
into this radial slice, computed at the position of
the tail of each arrow.  The Rosseland mean and effective photospheres of
the horizontally averaged structure at this time are indicated by the
upper and lower dashed lines, respectively.
\label{fig:rhoyzupperpgas}}
\end{figure}

\begin{figure}
\epsscale{1.0}
\plotone{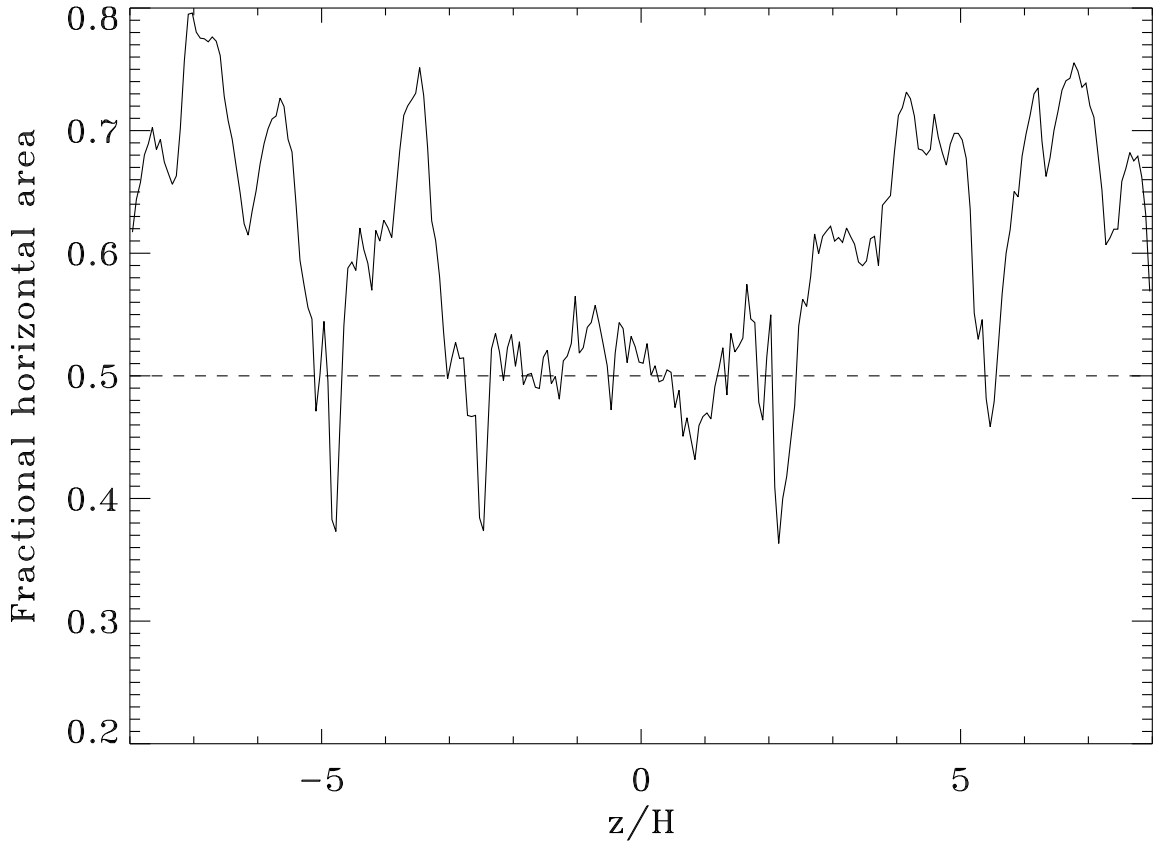}
\caption{Same as Fig.~13 except for the gas pressure dominated simulation
of \citet{hir06} at time 60 orbits.
\label{fig:pgastensionrhostats}}
\end{figure}

\begin{figure}
\epsscale{1.0}
\plottwo{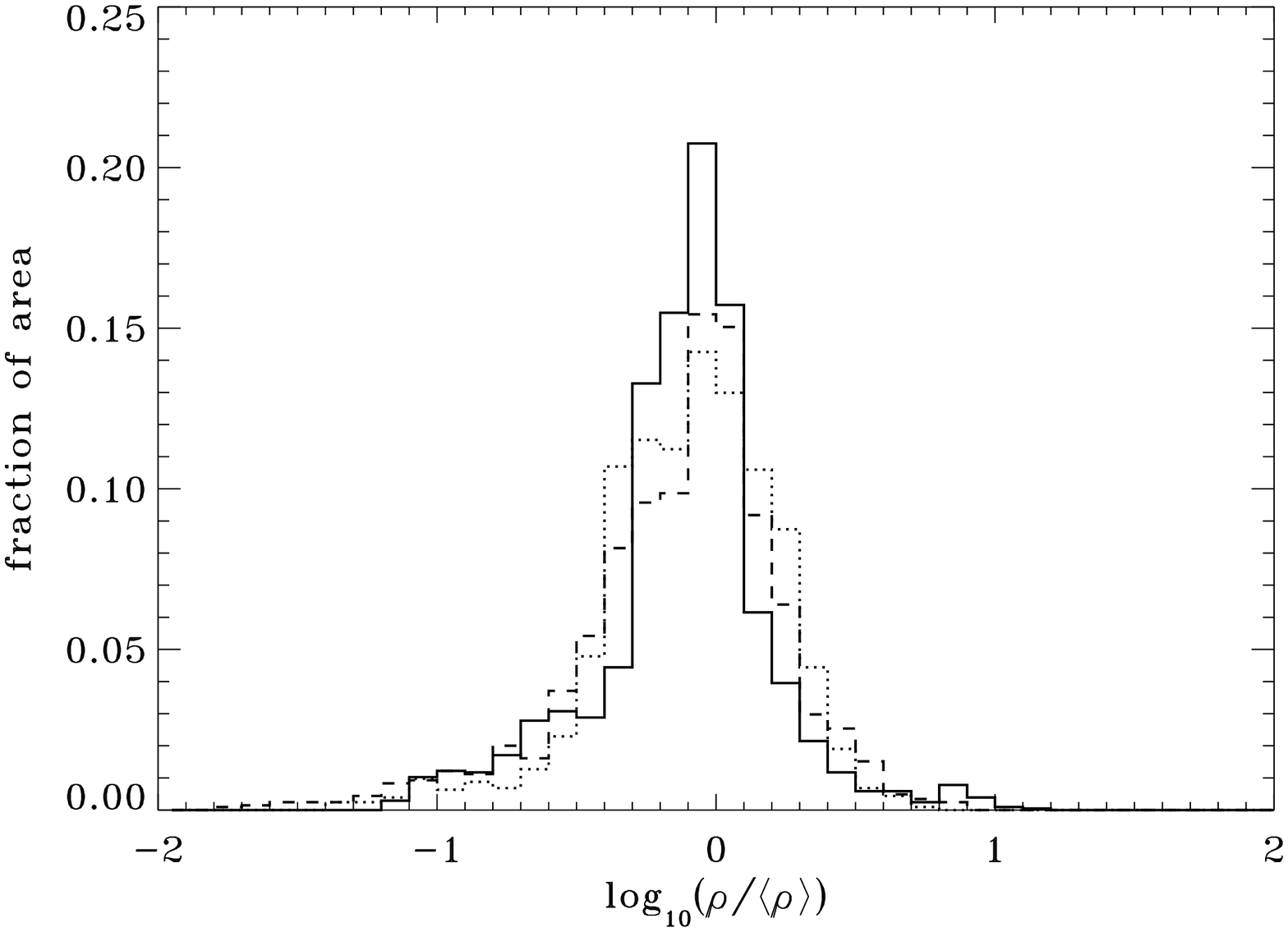}{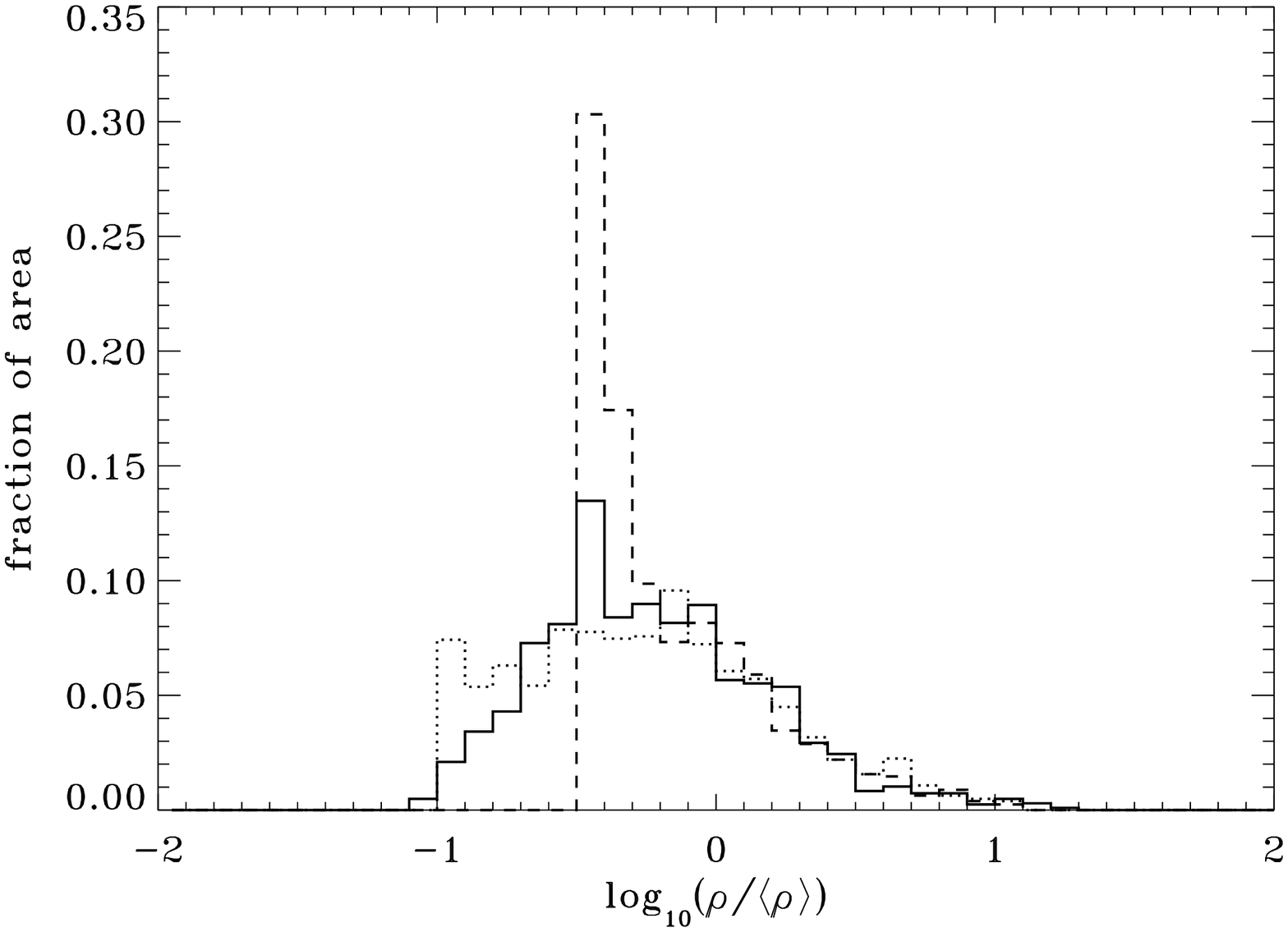}
\caption{Distribution of densities, scaled with the horizontally averaged
density, at the heights of the upper effective (left) and Rosseland mean (right)
photospheres at 90 orbits
(solid histograms) and 150 orbits (dashed histograms).  For comparison, the
dotted histograms show the corresponding density distributions
in the gas pressure dominated simulation of \cite{hir06} at 60 orbits.
\label{fig:rhofluct}}
\end{figure}

\begin{figure}
\epsscale{1.0}
\plotone{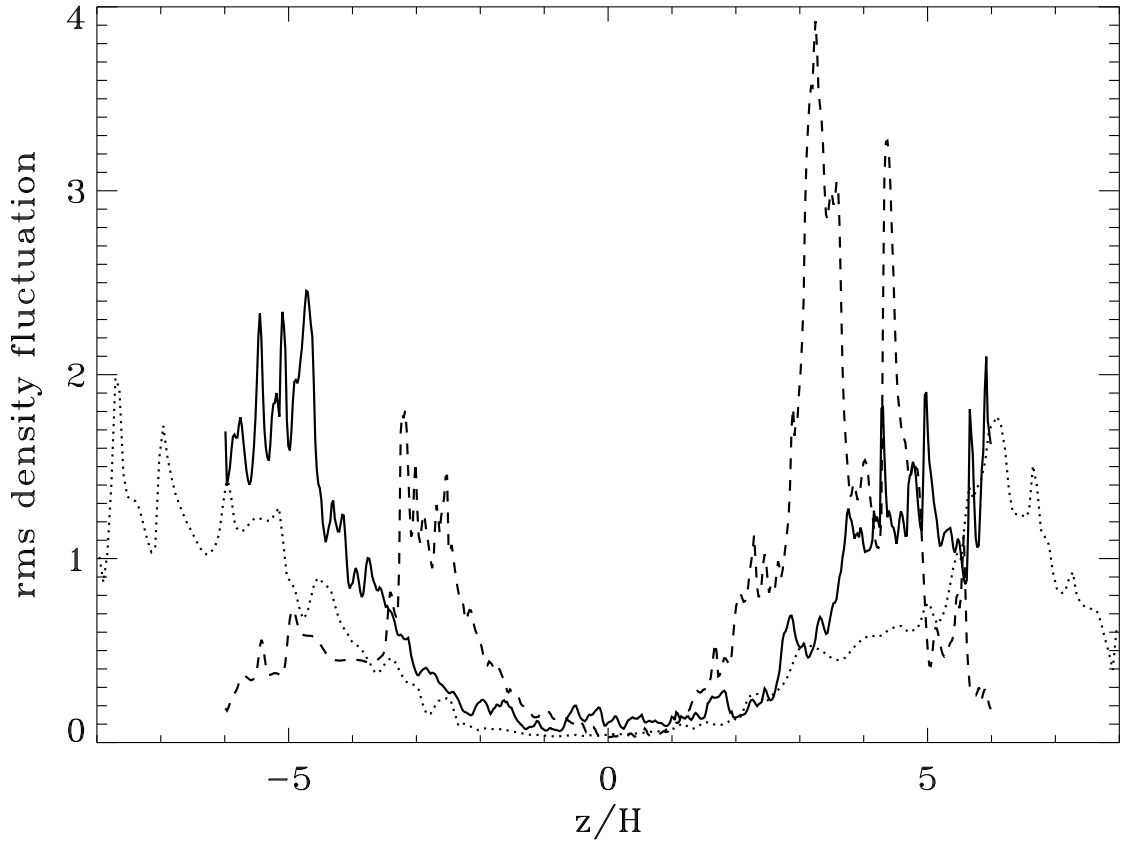}
\caption{Root mean square fractional fluctuation of density as a function
of height at 90 orbits (solid) and 150 orbits (dashed).  For comparison,
the dotted curve shows the same thing in the gas pressure dominated
simulation of \citet{hir06} at 60 orbits.
\label{fig:rhormshave}}
\end{figure}

\begin{figure}
\epsscale{1.0}
\plotone{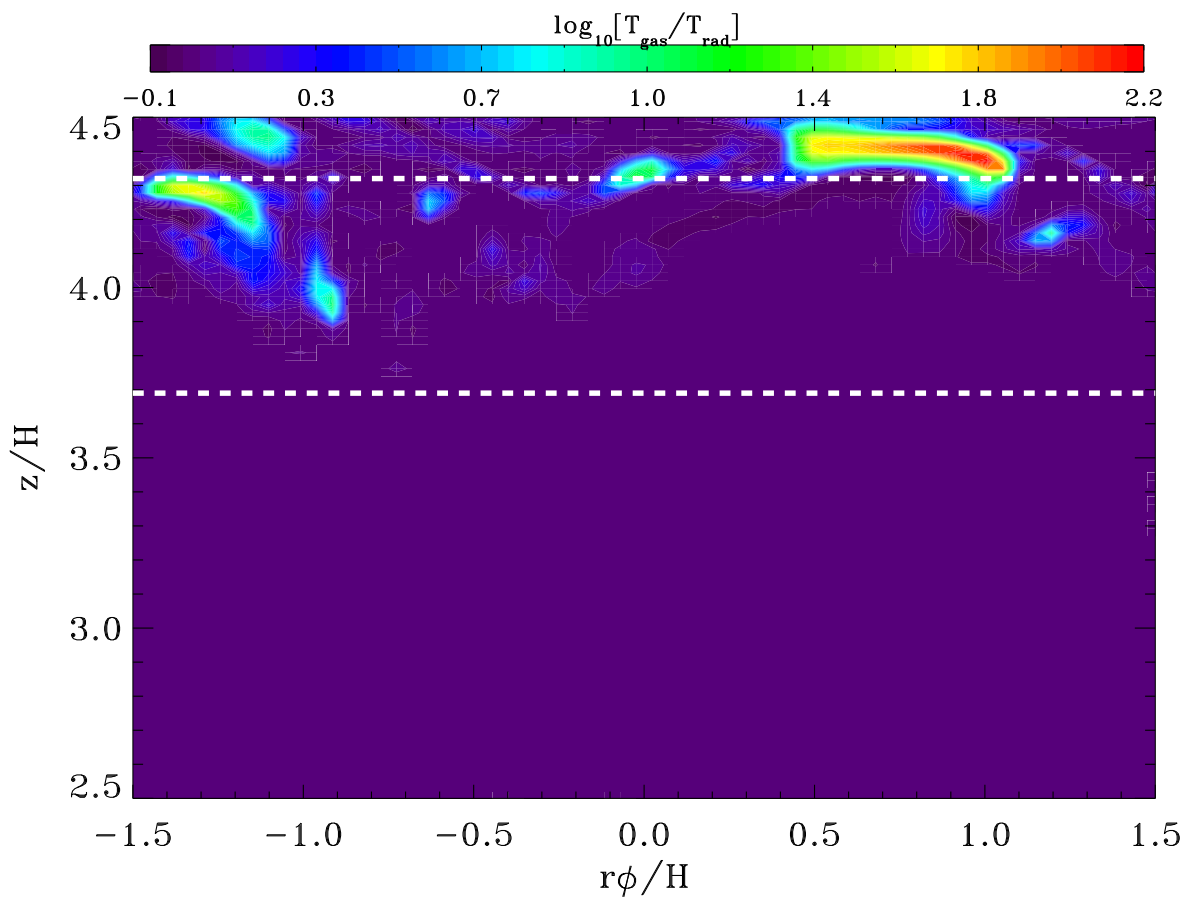}
\caption{Ratio of gas to radiation temperature near the upper photosphere
in a fixed radial slice near the middle of the box at t=90 orbits.  (The
radial slice is identical to that chosen in Fig. \ref{fig:rhoyzupper090}.)
Horizontal dashed lines indicate the positions of the Rosseland mean
photosphere (upper) and effective photosphere (lower), computed from the
horizontally averaged structure.
\label{fig:tgasotrad090}}
\end{figure}

\clearpage

\begin{figure}
\epsscale{1.0}
\plotone{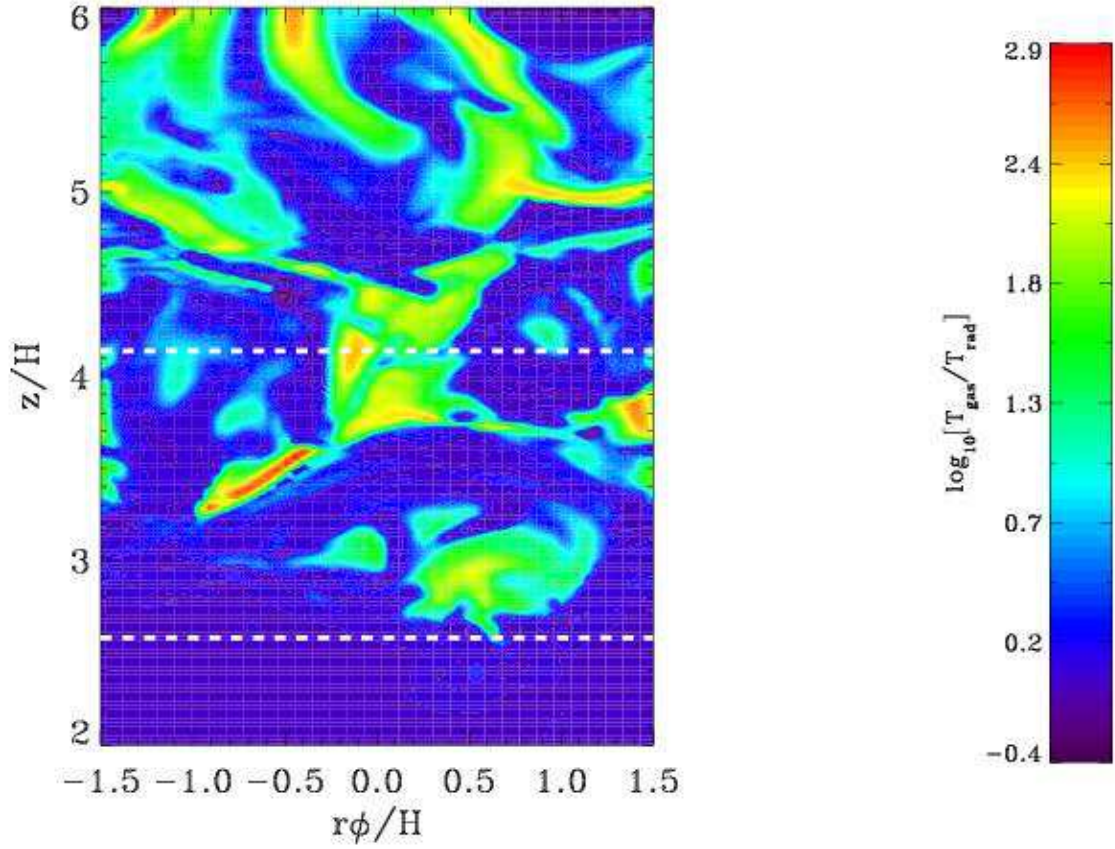}
\caption{Same as Fig. \ref{fig:tgasotrad090} except at t=150 orbits.
(The corresponding density structure is shown in Fig.~\ref{fig:rhoyzupper150}.)
\label{fig:tgasotrad150}}
\end{figure}

\end{document}